\documentclass[11pt]{article}
\usepackage{amsfonts}
\usepackage{eurosym}
\usepackage{amssymb}
\usepackage{adjustbox}
\usepackage{amsmath}
\usepackage{epsfig}
\usepackage{bigstrut}
\usepackage[flushleft]{threeparttable}
\usepackage{adjustbox}
\usepackage{xcolor}
\usepackage{verbatim}
\usepackage{natbib}
\usepackage[colorlinks=true,citecolor=blue,linkcolor=blue,urlcolor=blue,pdfpagemode=None]{hyperref}
\usepackage[caption=false]{subfig}
\usepackage{lipsum}

\newcounter{theorem}

\newtheorem{observation}{\sc Observation}

\newcommand{\SN}{{\rm I\kern-.23em N}}

\begin{document}
\title{The effect of ambiguity in strategic environments: an experiment\thanks{%
We wish to thank Maja Adena, Ciril Bosch-Rosa, Dirk Engelmann, Aniol Llorente-Saguer, Frederic Moisan, Salvatore Nunnari, and Marie-Claire Villeval, as well as
participants to seminars at Burgundy Business School, ECOBAS (U. Vigo), VIBES, Universidad Autónoma de Madrid, Bocconi, Lyon Business School, and Berlin Behavioral Seminar. We gratefully acknowledge the
financial help of the European Union's Horizon 2020 research and innovation
programme under the Marie Sklodowska-Curie grant agreement No 891124; MINECO-FEDER (PID2021-126892NB-I00,  PID2019-108718GB-I00); Basque Government IT1461-22.}}

\author{P. Brañas-Garza\thanks{Loyola Behavioral Lab. E-mail: branasgarza@gmail.com}, A. Cabrales\thanks{Universidad Carlos III de Madrid. E-mail: antonio.cabrales@uc3m.es}, M.P. Espinosa\thanks{University of the Basque Country UPV/EHU. E-mail: mariapaz.espinosa@ehu.eus}  \& D. Jorrat\thanks{Loyola Behavioral Lab. E-mail: dajorrat@gmail.com}} 
\date{\today} 

\maketitle

\begin{abstract}
    We experimentally study a game in which success requires a sufficient total contribution by members of a group. There are significant uncertainties surrounding the chance and the total effort required for success. A theoretical model with max-min preferences towards ambiguity predicts higher contributions under ambiguity than under risk. However, in a large representative sample of the Spanish population (1,500 participants) we find that the ATE of ambiguity on contributions is zero. The main significant interaction with the personal characteristics of the participants is with risk attitudes, and it increases contributions. This suggests that policymakers concerned with ambiguous problems (like climate change) do not need to worry excessively about ambiguity.
\end{abstract}

\newpage

\section{Introduction}

There is no doubt that the world will warm considerably in the next century. However, the uncertainty surrounding the precise magnitudes about the process, both the extent of warming and the effects of abatement, is also very large and may affect our willingness to contribute towards the effort. This problem is further compounded by the strategic interactions between abatement efforts (I may want to contribute more, if I think others are doing the same, or vice versa). There are many problems that share these two characteristics of climate change; first, humanity has not confronted them before, and thus there are fundamental uncertainties and, second, the contributions of various actors induce strategic interactions. These two aspects complicate decision-making considerably for the parties involved. We want to contribute to this debate by studying theoretically and experimentally how citizens react to scenarios that are conceptually similar.

With respect to the existing literature on decision making under uncertainty, we contribute along several dimensions. First our experimental participants make decisions in an ambiguous
environment where in addition there is strategic interaction and, because
of equilibrium multiplicity, an additional layer of
\textquotedblleft strategic uncertainty\textquotedblright. As we point out in the literature review, research tends to concentrate on either uncertainty in the environment (with no strategic interaction) or uncertainty about the actions of other players. Perhaps more importantly, this combination of uncertainties is also extremely relevant for our
intended applications. To the extent that climate change entails
\textquotedblleft tipping points\textquotedblright\ \citep{lenton2011early,lontzek2015stochastic}, this creates the possibility that decisions on abatement efforts present a
structure that is similar to a public good mechanism with a threshold or
\textquotedblleft provision\textquotedblright\ point \citep{isaac1989assurance,rondeau1999voluntary}\footnote{Another problem with the same structure is the accumulation of space debris, which will eventually reach a tipping point, known as the Kessler Syndrome, that would prevent any space activity \citep{adilov2018economic}.}.

Another crucial difference is that we run our experiment with a large
representative sample of the population of a mid-size country (Spain), and
we elicit many relevant characteristics about personal and political
preferences, as well as their socioeconomic background. In this way we can elicit
a granular picture of the reactions towards the problem which, although unusual in
many economic experiments, is very important for the situation at hand, a truly global phenomenon.

There have been numerous variations on the modelling of this problem but our
focus would be in ascertaining the effect of large uncertainty on citizens,
to then extract implications for regulators. These are non-trivial to
predict. On the one hand, a standard tool for the analysis of uncertainty,
maxmin preferences (arising from the pioneering work of \cite{gilboa1989maxmin}, 
tend to make decision-making more conservative. But for many problems, like our motivating example of
climate change, the implications are unclear a priori. It is true that by
focusing on the \textquotedblleft worst\textquotedblright\ possible prior,
the utility of doing nothing is very low. But pessimism can also affect
\textquotedblleft active\textquotedblright\ policies. Even worse than not
doing anything and suffering the consequences, is spending a lot in
abatement and then suffering those same (or very similar) impacts. This is further
compounded by the fact that the pathways between the actions the regulators can take
and climate change are not straightforward. Moreover, the costs are close in time
and there is not much uncertainty about them, whereas the benefits are
far in the future and much more uncertain.

Because of these difficulties, we take a theoretical and experimental approach to the
problem. We construct and analyze a game where the players face a
large uncertainty which can be mitigated with effort, 
but the efforts of several players are necessary for mitigation.
Then we run an experiment where a sample of the Spanish population, representative in terms of gender, 
age and education levels, is presented with vignettes about this interactive decision problem. 
The experiment was structured as follows.

Every participant in the experiment was placed in a group of 5 people, all of
whom had an endowment of money. There was a chance that the money of the
group would disappear (it could be stolen), but the members could make a voluntary
contribution to a fund (to improve the safety of the safe) 
that, if sufficiently large, mitigated that uncertainty.

The experiment was run in June 2021. 
It had four treatments which represented uncertainty or risk about
the two dimensions of the problem. On the one hand, the probability that the
money disappeared with/without the investment. Under risk, one out of five
times the money was lost if a large enough investment was made, and four out
of five if the investment was not large enough. Under uncertainty, the money
was lost at most two out of five times with enough investment and at least
three out of five times without investment. This parameter (the probability of total loss) can be thought of
as the climate consequences of undertaking abatement measures or not.

The other relevant parameter was the amount of investment necessary to prevent
the money from being lost. Under risk, the necessary amount could be either 5
or 10 euros, both with equal likelihood. Under uncertainty all that was known
was that the necessary amount was either 5 or 10. Again, the
analogy with climate change is that the amount of investment necessary to avoid
catastrophic consequences is uncertain.

In our experiment ambiguity is not salient, the framing is completely neutral about it. 
The fact that participants do not know the probabilities is never made explicit. 
For example, subjects are informed that at least 4 out of 5 times there will be total loss.
This implies that they do not know the probability for sure, but we do not emphasize that they do not know it.
In the case of ambiguity in the abatement costs, subjects are informed that the cost is in a given interval.

In addition to the vignettes, the participants were tested with standard
measures to evaluate their attitudes to risk, uncertainty, distributional
preferences, and time preferences. They also answered a socioeconomic questionnaire that
included, in addition to standard variables, all the questions from the Eurobarometer 
related to attitudes towards climate change and environmental problems.\footnote{https://ec.europa.eu/clima/citizens/citizen-support-climate-action\_en}

To generate hypotheses for the experiment, we constructed a model for the
behaviour of experimental participants in the spirit of \cite{gilboa1989maxmin} where agents are endowed with maxmin preferences. The
predictions are very clear, provided there are at least some participants who
are risk-loving. The contributions should be largest on average in the
treatment with uncertainty in the two dimensions. The following treatment in
terms of the amount of predicted contributions is the one where there is uncertainty
only in the probability of avoiding the damage, but risk on the threshold.
The other two treatments are harder to rank in terms of average
contributions, but they should be more polarized in the one for which there
is uncertainty about the threshold and risk about the probability of
avoiding the damage, than the one where there is risk in the two dimensions.
The \textquotedblleft pessimism\textquotedblright\ inherent in the maxmin
formulation of preferences is key to understanding the differences in
theoretical predictions and the larger expected contributions where
uncertainty dominates.

The empirical results are mostly in contrast with the predictions of our theoretical benchmark.
There were no differences in contributions between the treatments. Given the
really large sample size used for the experiment, 1500 people, the result is not likely
due to an absence of statistical power. We obtain a very
accurately estimated zero effect for all the treatments. 

A second important result is that contributions were significantly
smaller for individuals who are risk averse.  This can be seen as puzzling. 
A best-response by a risk averse person to fears of others' low contributions could be to contribute more 
to ensure reaching the threshold. But we also show that risk averse participants believe that others contribute little. 
Thus, lower contributions can arise from false consensus \citep{engelmann2000false} 
or projection bias \citep{cason2014misconceptions}. 
 This has been shown to be important in other public good games \citep{smith2015modeling}.

Our third and last result regards interactions. The main one that shows significance is related to risk aversion. 
Risk averse individuals who participated in the treatment with risk in the two dimensions contributed less 
than risk averse individuals in all other treatments. 
The next level of contributions (among the risk averse) came from participants in the treatment where there is uncertainty
only in the probability of avoiding the damage, but risk on the threshold. 
Risk averse participants in the other two treatments were the ones who contribute the most, 
and they were not significantly different between themselves. That goes partly in the direction of our 
theoretical benchmark predictions. 

When we interacted our treatments with other variables that could be expected to yield heterogeneous effects, 
such as gender, mathematical ability or reflectiveness (the score of the CRT test), we did not find any significant effects. 
This latter result is important. People who score high on CRT tend to think through the questions and reflect seriously 
on them (this is literally how one scores high on CRT). Thus, it is unlikely that the lack of impact of our treatments occur 
simply because participants are not paying attention to the problem. One possible interpretation of this result points an important direction for future research. We conjecture that for many ``ordinary'' people, their perception of uncertainty 
is not very precise even if they are given precise probabilities. Thus, when given either intervals or point 
estimates, their cognitive mappings could be very similar, in line with the theory of \cite{khaw2021cognitive}.

To understand the mechanisms behind these results, we examine the effect of
the treatment and other variables on beliefs about others' contributions. We
find that the treatments do not affect how much individuals believe others
are going to contribute. However, risk averse individuals have a
pessimistic belief about the contribution of others.

The main policy implication of the paper is that policymakers concerned with problems that are naturally ambiguous, such as climate change, should not be overly concerned with the ambiguity dimension. It would either not matter, or, for some people, it would increase contributions. On the other hand, since risk averse people contribute less, lowering the perception of risk would be useful.
This could be done by communicating that there are adverse consequences about which there is no uncertainty and they cannot be avoided unless decisive action is taken quickly.

\section{Related literature}

\cite{quiggin2005precautionary} points out that an alternative course of action proposed in
uncertain environments is a \textquotedblleft precautionary
principle.\textquotedblright\ This principle states that when reverting the
impacts of a damage is more difficult than preventing it, one must shift the
benefit of proof in favor of those proposing preventive action. In the case
of climate change, that means mitigation. But as Quiggin also indicates
forcefully \textquotedblleft in the discussion of the precautionary
principle, there has been only occasional reference to the literature on the
theory of choice under uncertainty, a literature that spans economics,
psychology and statistical decision theory. The absence of any formal
framework for discussion has contributed to the confused nature of the
debate.\textquotedblright

In situations such as those we use for motivation, individual decision-makers often
cannot agree on the prior probability distributions. One way
to deal with this problem, pioneered by \cite{gilboa1989maxmin} is to dispense with the assumption that agents have a single probability distribution to make decisions and instead consider multiple priors. Then, for each agent,
the different actions are ordered by focusing on the
distribution that yields the worst expected utility, and then choosing the
one giving the maximum utility. This procedure is called maxmin expected
utility with multiple priors. There are many generalizations of this
process. Possibly the most developed one is \cite{maccheroni2006ambiguity}. For a survey of models of ambiguity aversion, see \cite{machina2014ambiguity}.

There is an important literature applying ambiguity in strategic contexts theoretically. In \cite{eichberger2002strategic} strategic ambiguity has opposite effects in games of strategic complements and substitutes; for strategic substitutes, increasing the level of ambiguity changes equilibrium strategies in an ex-post Pareto improving direction, while for strategic complements, an increase in ambiguity has the opposite effect. 
\cite{calford2021mixed} studies ambiguity in games with mixed-strategy equilibria.

In terms of experiments, an important one for us is \cite{berger2020policymakers}, who conducted one among participants in the COP21 (UN Climate Conference in Paris) and showed they were
on average ambiguity averse. While this is an interesting investigation
about the preferences of policymakers, their participants are far from being
representative of the world population. In most countries the population at
large has an important role in policy determination via the democratic
process. Moreover, the participants took part in an
individual choice experiment and strategic considerations were absent.

There are also many papers studying ambiguity aversion in games. For example, 
\cite{ivanov2011attitudes} uses games to identify whether someone is ambiguity loving, neutral 
or averse and finds out the proportions are  32\%, 46\%, and 22\% respectively. 
\cite{kelsey2015experimental} show that in coordination games, players tend to avoid ambiguous strategies. 
\cite{kelsey2017dragon} repeats this exercise for best-shot and weakest link games. 
\cite{kelsey2018strategic} test experimentally the effect of strategic ambiguity; they find it has a larger effect 
on games than on individual decision making problems, and opposite effects in games of strategic substitutes and complements. 
These three papers concentrate on ambiguity in the distribution of the other players' actions, 
rather than on uncertainty about the consequences of (profiles of) actions.

Previous literature on ambiguity in public good games with an interior solution has also focused on strategic uncertainty 
(i.e. ambiguity about what other players will do); a main result is that a larger ambiguity puts a higher weight on lower
contributions by the opponents and therefore increases contributions (e.g. \cite{eichberger2008granny,di2011kindness}).
\cite{dannenberg2011coordination} find that threshold uncertainty has a negative effect on coordination success.

Our contribution to this experimental literature on ambiguity in strategic environments is to analyze the effect 
of ambiguity that is inherent to the environment.\footnote{There are of course other reasons why one can expect
strategic complementarities, such as the fact that many innovations have
important spillovers between themselves \cite{chalioti2019spillover}.}  
We also use a representative sample of a population, thus making the external validity issues less important in our case.

\section{Experimental design}

\begin{figure}
    \centering
    \includegraphics[width=11cm]{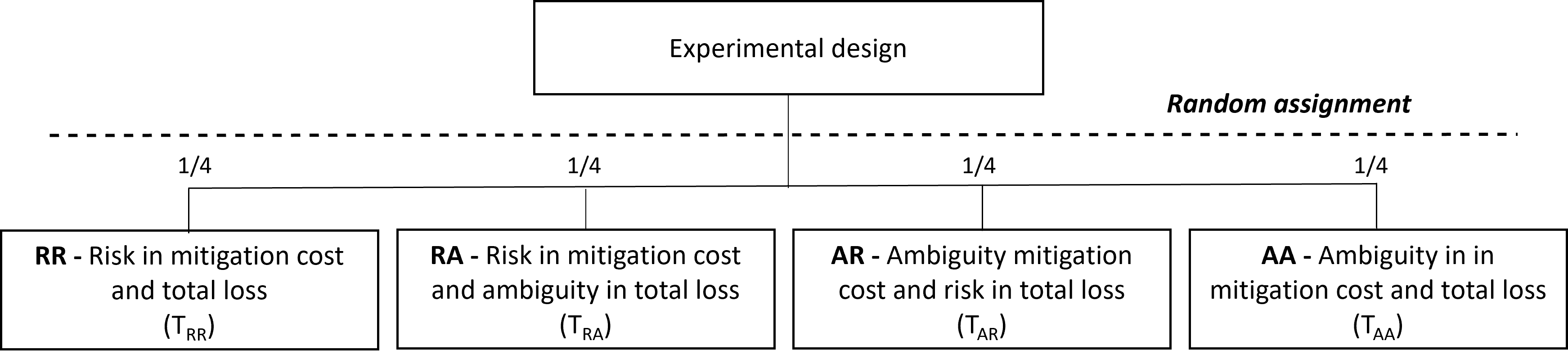}
    \caption{Randomisation process}
    \label{fig:design}
\end{figure}

In order to test our hypotheses (described in detail in subsection \ref{hypotheses} and based on the theoretical benchmark from section \ref{theory}) we designed an experiment with $2\times 2$ treatments (see Instructions\footnote{Instructions in English are available here: http://hdl.handle.net/20.500.12412/3402.\label{Instructions}}): there was either risk or ambiguity in the threshold and the probability of averting the disaster. We think this is an important characteristic of our design. In reality, uncertainty can impinge on various variables, and as we will see in the theoretical benchmark, the effect is not equivalent depending on the variable on which the uncertainty impinges. This is relevant practically because it could guide whether we want to invest in reducing the uncertainty along the different margins. Figure \ref{fig:design} summarizes the features of our design. 

The baseline treatment was (\textbf{RR}), with a precise probability distribution on both the probability 
of total loss and the mitigation cost. The other treatments had ambiguity in either the probability of total loss, 
but not in the mitigation cost (\textbf{AR}), ambiguity in the mitigation cost but not in the probability 
of total loss (\textbf{RA}), or ambiguity in both (\textbf{AA}).

We described the problem as that of a group of 5 people, each of whom had an endowment of 5 Euros in a safe. But a thief might
steal all the money in the safe, in which case all group members would lose all their money. The risk of a total loss could be
mitigated, at a cost. If the group members (simultaneously and privately) put enough money in a common pot, the theft was prevented
with 0.9 probability (under risk) and with ``at least'' 0.8 probability (under ambiguity). The collective contribution threshold was
either 5 Euros or 10 Euros with equal probability (under risk) and either 5 Euros or 10 Euros (under ambiguity, with no mention of a
probability). This situation shares features with our motivating climate change problem (the need to contribute under uncertainty to
avoid the disaster, the uncertainties about abatement efforts and the strategic interaction); but it was more concrete 
and adequate for the relatively small amounts that were at stake. 

\begin{figure}
    \centering
    \includegraphics[width=12cm]{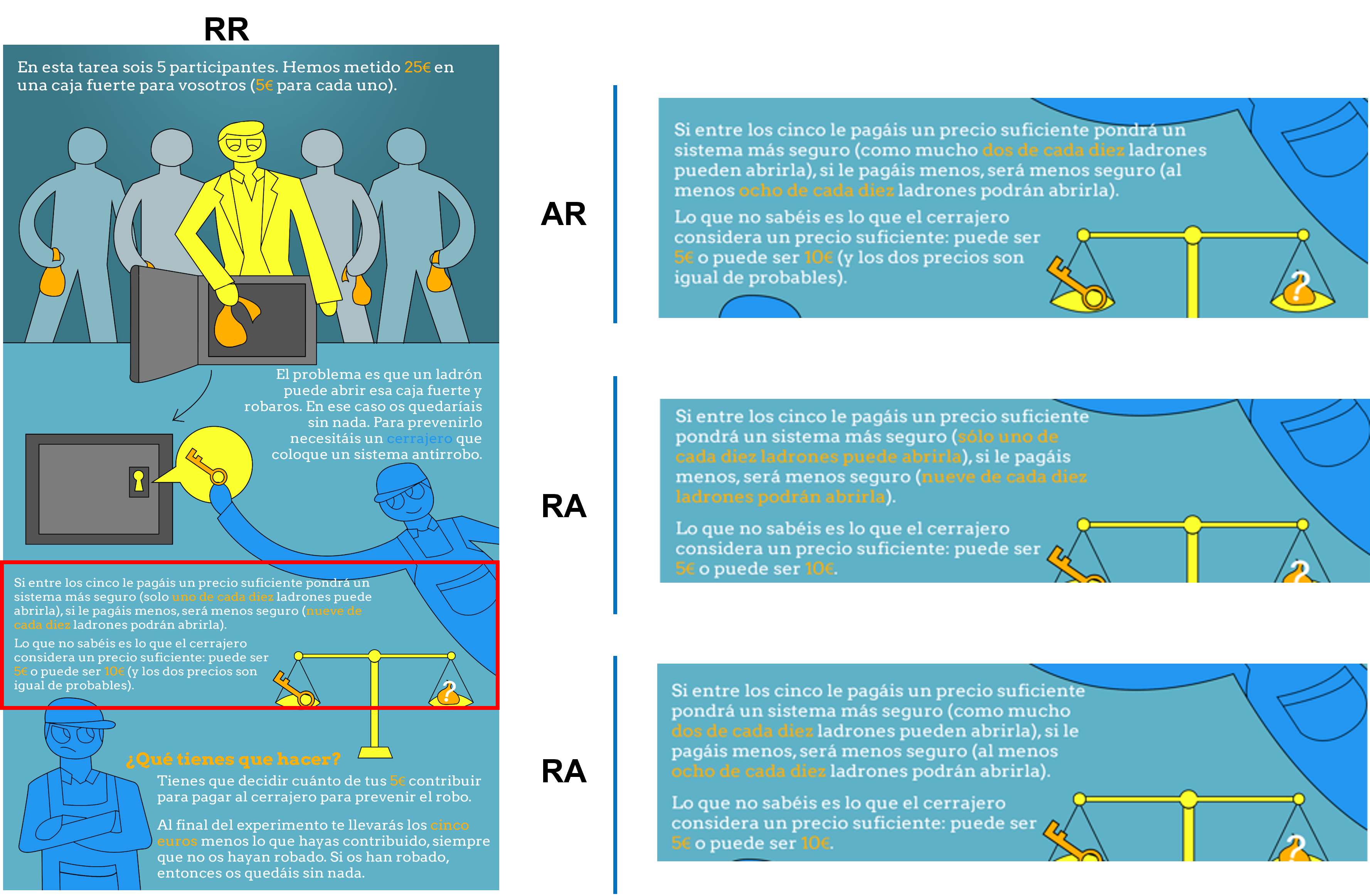}
    \caption{Treatments}
    \label{fig:treatments}
\end{figure}

Since the sample included people from many different social backgrounds, we went for a simple design, where we did not even mention
probabilities but frequencies. That is, we did not say ``0.9 probability'', but rather ''9 out of 10". Also, we went for a careful
graphical design, that we pre-tested on a sample of students and non-students for ease of comprehension and simplicity. 
Figure \ref{fig:treatments} shows the instructions for \textbf{RR} on the left, where the only thing that changed in the other
treatments was the red box, which is then shown for the other three treatments on the right panel.

\subsection{Covariates}

There were several variables that might plausibly interact with the treatments. Two quite prominent ones were \emph{risk aversion} and
\emph{ambiguity aversion}. Clearly, ambiguity averse participants may be more affected by the ambiguity treatments than the people who are
not, and this may translate into their willingness to contribute. The connection with \emph{risk aversion} was less obvious in
terms of interactions with the treatments, but it seemed an important control in this environment. 

\begin{figure}[ht!]
\centering
 \begin{minipage}[m]{0.45\textwidth}
\includegraphics[scale=0.05]{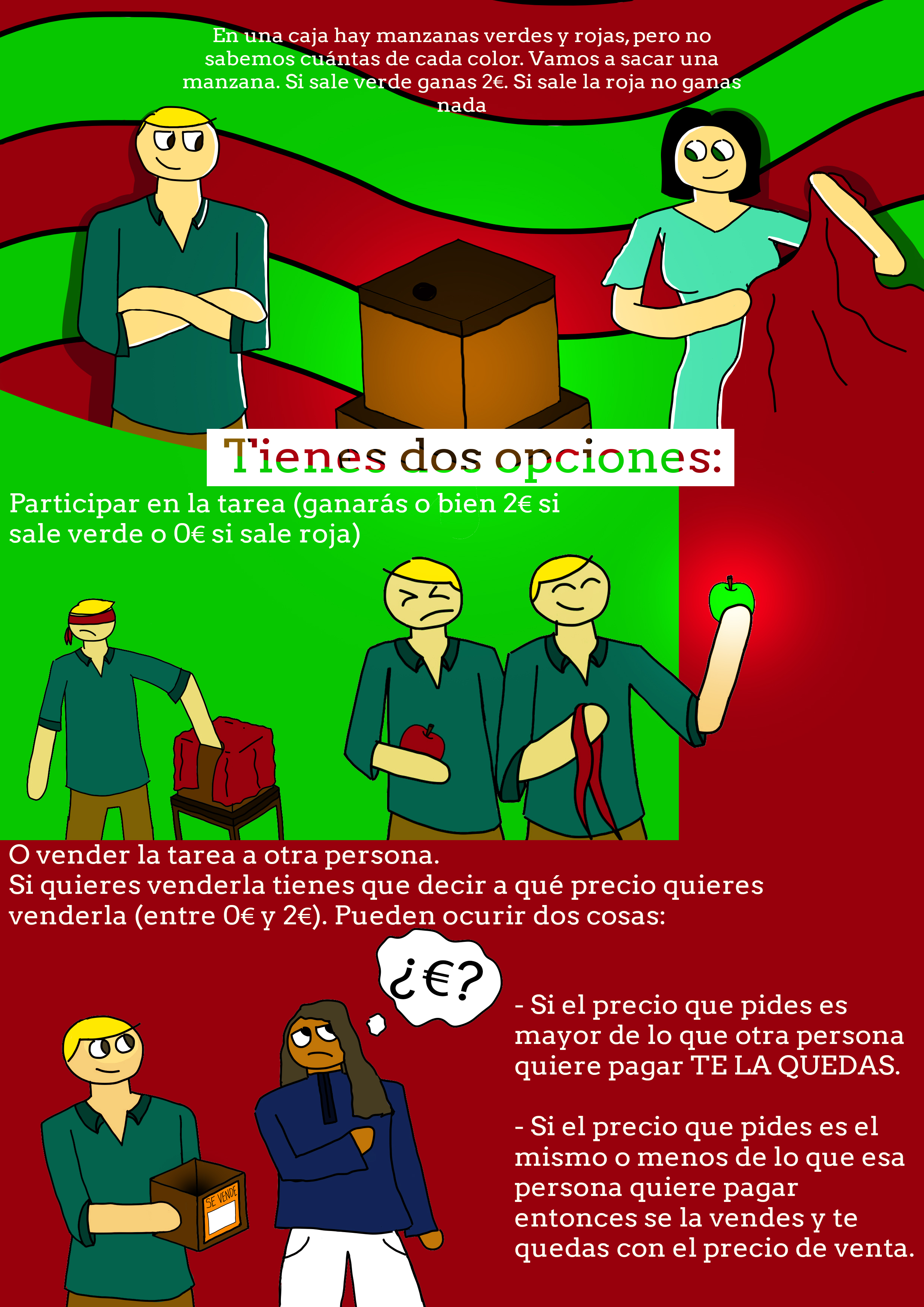}
\caption{Ambiguity aversion}\label{ambiguity}
 \end{minipage}
 \quad
 \begin{minipage}[m]{0.45\textwidth}
\includegraphics[scale=0.05]{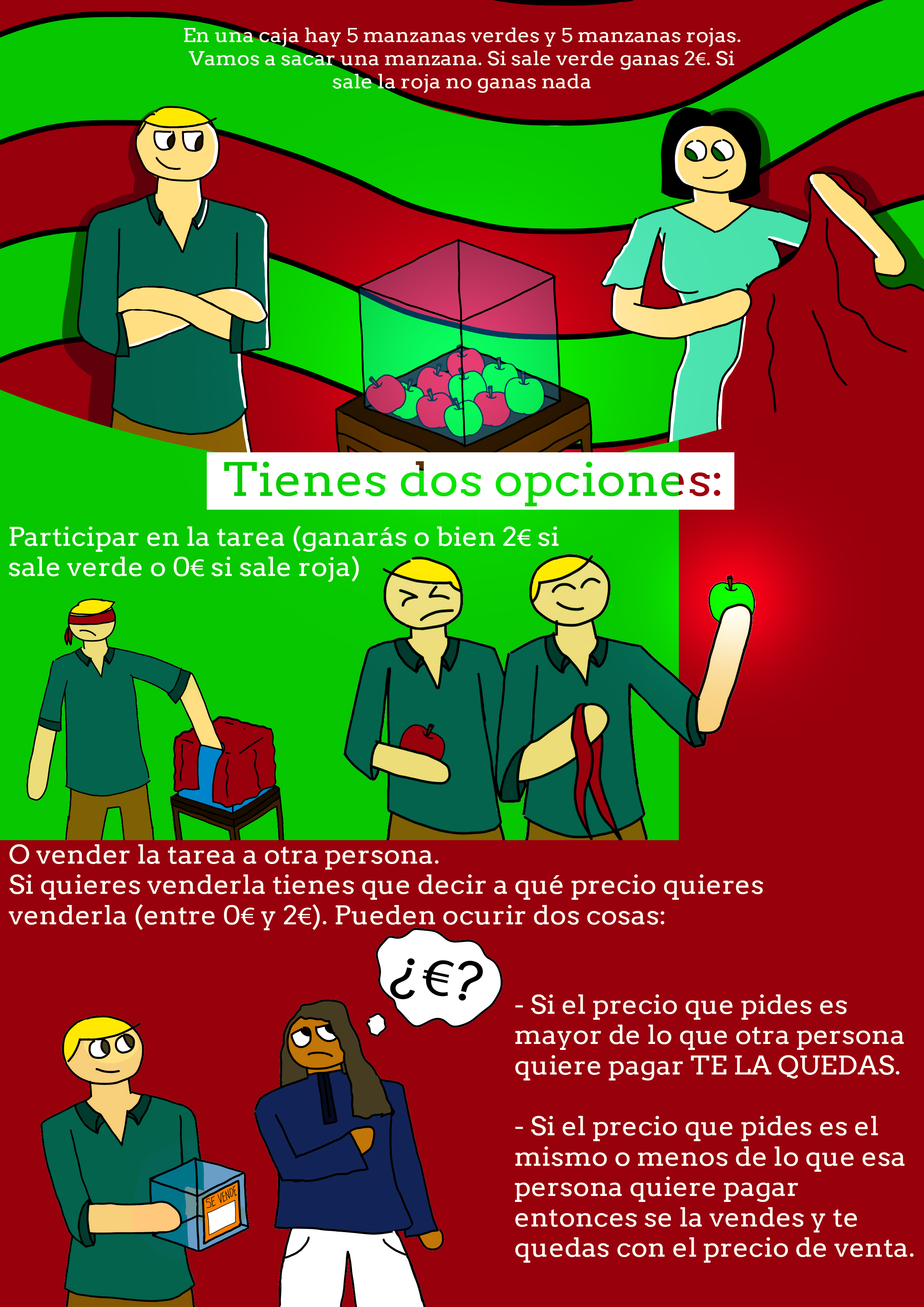}
\caption{Risk aversion}\label{risk}
 \end{minipage}
 \end{figure}

As with the main treatments, given the nature of the subject pool, we chose a simple way to elicit the preferences. In both cases,
we elicited the willingness to pay for avoiding a risky or ambiguous task. For risk, we presented an urn with 5 green 
and 5 red apples. The participant received 2 Euros if a green apple was extracted and 0 otherwise, if she decided to take this task.
But she could sell the task to another person, and we elicited the value at which she would do it. The ambiguity task was the same,
except that we did not give the number of red and green apples. Figure \ref{risk} depicts both tasks, ambiguity on the left 
panel and risk on the right panel. 

{It is worth mentioning that our measure of ambiguity is based on both tasks, and is defined for each subject as the difference between the selling price of the risk lottery and the selling price of the ambiguity lottery. Negative values of this variable mean that the subject preferred the ambiguous lottery over the uncertain one (i.e. they were ambiguity loving), while positive values mean that the uncertain lottery is preferred over the ambiguous one (i.e. they were ambiguity averse). For those participants who answered the same price on both tasks or kept both lotteries, this difference equals zero (i.e., they were indifferent to uncertainty and ambiguity).}

\begin{figure}[ht!]
\centering
 \begin{minipage}[m]{0.45\textwidth}
\includegraphics[scale=0.04]{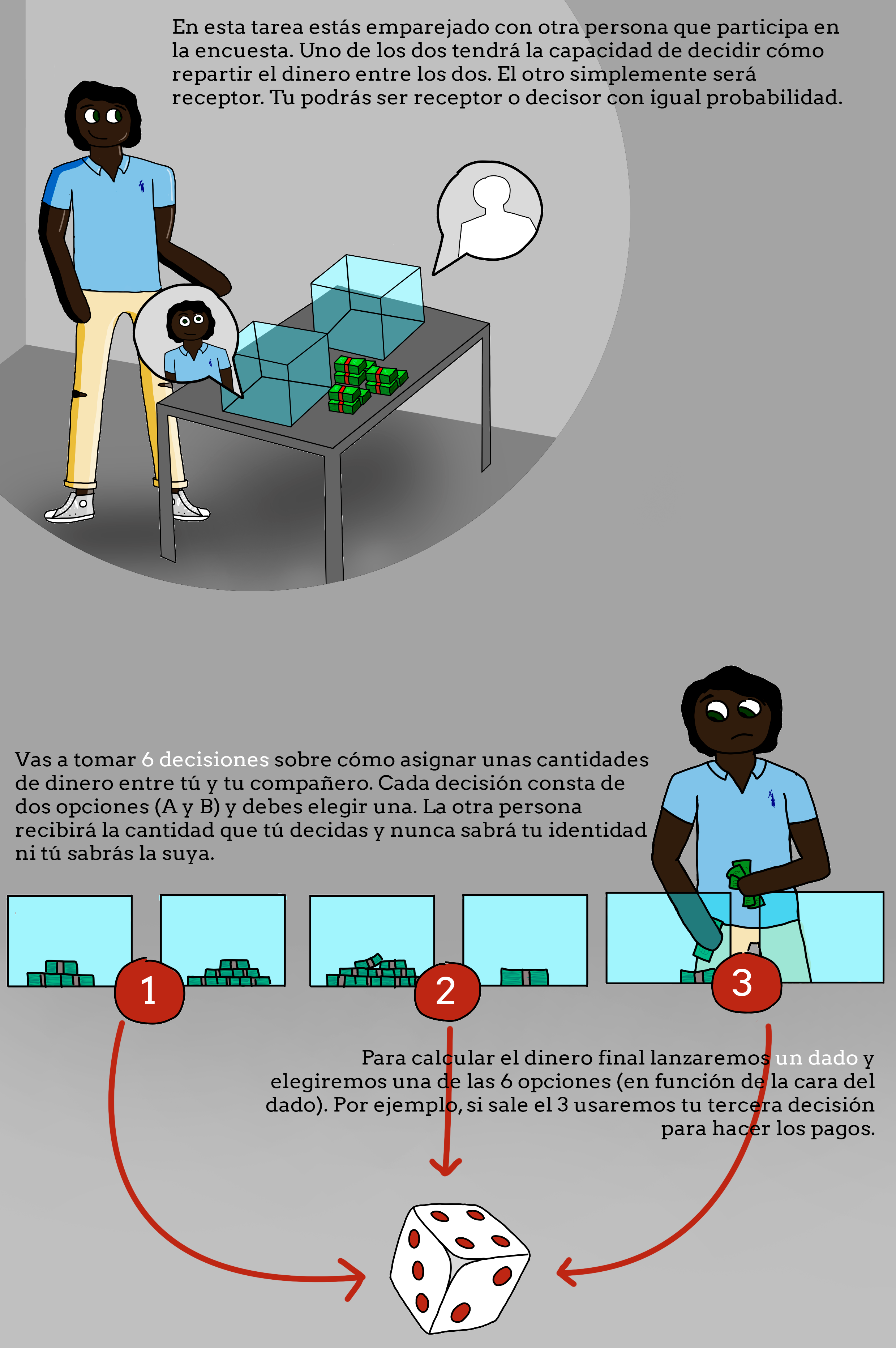}
\caption{Dictator game}\label{dictator}
 \end{minipage}
 \quad
 \begin{minipage}[m]{0.45\textwidth}
\includegraphics[scale=0.05]{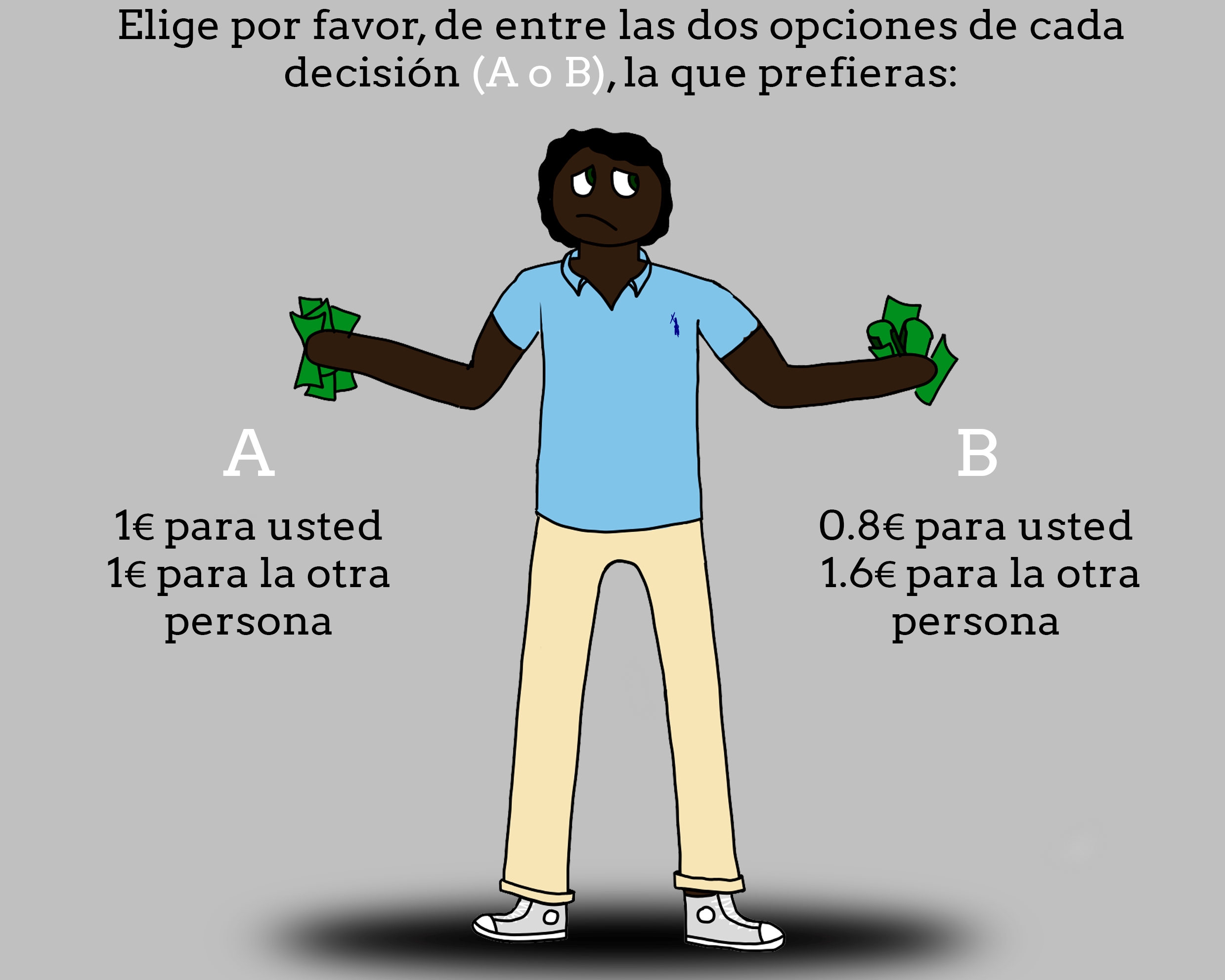}

 \end{minipage}
 \end{figure}

Another obviously important control variable in a game with contributions is social preferences. Following
\cite{corgnet2015cognitive}, we measured them with 6 dictator games, out of which one was randomly chosen for payment. 
The payoffs are as follows. In dictator game 1, the choice is between the pair (1, 1) and the pair (0.8, 1.6), 
where the first payoff is always for the dictator and the second for the other player. In game 2, the choice is between (1, 1)
and (1.2, 0.4), in game 3 between (1, 1) and (1, 0.6), in game 4 between (1, 1) and (1.6, 0.4), in game 5 between (1, 1) 
and (1, 1.8), and in game 6 between (1, 1) and (1.1, 1.9). The six games were presented in a graph where the top part was the left
panel of Figure \ref{dictator} and the bottom part was the right panel of Figure \ref{dictator} where the numbers 
change depending on the game.
    
\subsection{Additional questions and sociodemographics}
We also elicited additional information, of potential relevance. The participants took a Cognitive Reflection Test, 
and a simple math comprehension test, where they had to do some simple divisions and think about compound interest. 
The rationale was that perhaps more reflective people, or those with better mathematical skills would behave differently. 
They also took a  (non-incentivized) time preferences test.\footnote{A recent study shows that hypothetical and incentivized time preferences are mostly the same, see \cite{branastime}.}

In addition to the preferences elicitation tasks, we had a socio-demographic questionnaire with 34 items. Among other questions, 
we asked about gender, education, profession, income, political preferences, social attitudes, social origin, generalized trust, 
trust in government and institutions, perceptions about the most urgent social problems, and attitudes toward climate change 
(these latter were the entire set of questions on the topic from the Eurobarometer). 
See Instructions for details (foonote \ref{Instructions}).

\section{Theoretical benchmark}\label{theory}
There are $n$ players and each has an initial endowment of 5. There is a probability that they lose everything (e.g. climate disaster),
but they can implement mitigation measures that are costly. The mitigation measures have to be undertaken by the group and financed
through voluntary contributions. Denote by $c_{i}$ the contribution of each group member $i\in \{1,...,5\}$. When the probability 
that mitigation measures are effective is known, the objective function for each player is:

\begin{equation*}
\max_{c_{i}}U\left( c_{i},\sum\limits_{j\neq i}c_{j}\right) = u\left( 5-c_{i}\right)*
p\left( C\right)
\end{equation*}
where $u(x)$ is such that $u'(x)>0$ and $u(0)=0$,  $C=\sum\limits_{j=1}^{5}c_{j}=c_{i}+\sum\limits_{j\neq
i}c_{j}=c_{i}+C_{-i}$ and $p\left( C\right)$ is the probability that mitigation measures are effective and thus total loss 
is avoided, which is a function of the amount spent $C$. When mitigation measures are effective players keep what is left of the
endowment after the contribution.

In some of the treatments there was ambiguity either in the probability of total loss or the abatement costs so that $p(C)$ was not
known. In those cases we will assume a maxmin expected utility function, so that the objective of each player under ambiguity is: 

\begin{equation*}
\max_{c_{i}}U\left( c_{i},\sum\limits_{j\neq i}c_{j}\right)
=\max_{c_{i}}\left\{u\left( 5-c_{i}\right)* \min_{p} p\left(
c_{i}+C_{-i}\right) \right\}
\end{equation*}

The baseline distributions for $p$ are:
\begin{itemize}
\item [(a)]\label{prob} The cost of mitigation measures is either 5 with probability $0.5$ or 10 with probability $0.5$.

\item [(b)] If the cost of mitigation measures is reached through contributions, then the probability of total loss is reduced to
$0.1$. However, if the contributions fall short of the abatement cost, the probability of total loss is $0.9$. 
\end{itemize}

We focus on
symmetric undominated pure strategy Nash equilibria (no player's equilibrium strategy is weakly dominated) which are 
robust to any type of utility function $u(x)$ the subjects may have.
In all of the treatments the efficient solution is $C=10$. 

\subsection{Risk-Risk}

In this treatment both the distribution of the cost of mitigation measures and the distribution of the event of total 
loss are known, so from (a) and (b):

\begin{equation*}
p\left( C\right) =\left\{ 
\begin{array}{c}
0.1\text{ if }C<5 \\ 
0.1\ast \frac{1}{2}+0.9\ast \frac{1}{2}=0.5\text{ if }5\leq C<10 \\ 
0.9\text{ if }10\leq C<25%
\end{array}%
\right.
\end{equation*}%
Clearly, given that the threshold is either 5 or 10, in equilibrium $C$ can only be 0, 5 or 10. 
Note that depending on the subjects valuation $u$, we may find that in a Nash equilibrium total contributions are 0, 5 or 10. For
example, if subjects are risk neutral, it is easy to check that there are three symmetric equilibria with total contributions 
$C = 0$,  $5$ or $10$, respectively.
However, if we allow for some players to be sufficiently risk loving, the equilibria with positive contributions
could disappear (e.g. $u(x)=x^8$)

\begin{enumerate}
\item For an equilibrium with $C=0$ we need:
\begin{equation*}
U\left( 0,0\right) =0.1\ast u(5)>U\left( 5,0\right) =0.5\ast u(0)=0
\end{equation*}

which holds for any $u$.

\item For an equilibrium with $C=5$:
\begin{equation*}
U\left( 1,4\right) =0.5\ast u(4)>U\left( 0,4\right) =0.1\ast u(5)=
\end{equation*}

which holds as long as $u(5)<5*u(4)$.

\item For an equilibrium with $C=10$: 
\begin{equation*}
U\left( 2,8\right) =0.9\ast u(3)>U\left( 0,8\right) =0.5\ast u(5)
\end{equation*}

which holds as long as $u(5)<(9/5)*u(3)$
\end{enumerate}

\subsection{Risk-ambiguity}

In this treatment ambiguity affects the mitigation cost, so the probability distribution (a) in page 14 is replaced by:
\begin{itemize}
\item [(a')] The cost of mitigation measures is in $\{5, 10\}$ with unknown probabilities.
\end{itemize}

Thus, from (a') and (b), if we choose the lowest possible $p$ for each $c_{i}$ we get 
\begin{equation*}
p\left( C\right) =\left\{ 
\begin{array}{c}
0.1\text{ if }C<10 \\ 
0.9\text{ if }10\leq C<25%
\end{array}%
\right.
\end{equation*}

Since now the thresholds are 0 and 10, in equilibrium $C$ can only be $0$ or $10$.

\begin{enumerate}
\item For an equilibrium with $C=0$:
\begin{equation*}
U\left( 0,0\right) =0.1\ast u(5)>U\left( 5,0\right) =0.1\ast u(0)=0
\end{equation*}

which holds for any $u$.

\item For an equilibrium with $C=10$:
\begin{equation*}
U\left(2,8\right)=0.9*u(3)>U\left(0,8\right)=0.1*u(5)
\end{equation*}

which holds as long as $u(5)<9*u(3)$.
\end{enumerate}

\begin{observation}\label{ob1} 
Introducing ambiguity about the mitigation cost makes the equilibrium with $C=10$ easier to sustain ($u(5)<9*u(3)$  vs $ u(5)<(9/5)*u(3)$), which makes $C=10$ more likely and therefore suggests higher contributions under ambiguity. On the other hand, the contribution profile with $C=5$ is no longer an equilibrium with ambiguity about the mitigation cost.
\end{observation}

\subsection{Ambiguity-risk}

In this treatment ambiguity affects the chances of total loss. Thus, the probability distribution (b) in page 14 is replaced by (b'):
\begin{itemize}

\item [(b')] If the cost of mitigation measures is reached through contributions, then the probability of total loss is reduced to a value in [0, 0.2] with unknown distribution. However, if the contributions fall short of the abatement cost, the probability of total loss is a value in the interval [0.8, 1] with unknown distribution. 
\end{itemize}

Thus, from (a) and (b'), if we choose the lowest possible $p$ for each $c_{i}$ we get 
\begin{equation*}
p\left( C\right) =\left\{ 
\begin{array}{c}
0\text{ if }C<5 \\ 
0.8\ast \frac{1}{2}=0.4\text{ if }5\leq C<10 \\ 
0.8\text{ if }10\leq C<25%
\end{array}%
\right.
\end{equation*}

\begin{enumerate}
\item For an equilibrium with $C=0$:
\begin{equation*}
U\left( 0,0\right) =0*u(5)= 0=U\left( 5,0\right) =0.4*u(0)=0
\end{equation*}

so $C=0$ is a Nash equilibrium, but players are using weakly dominated strategies. If there is a chance that some other player has
contributed 1, then contributing zero is dominated by contributing 4.

\item For an equilibrium with $C=5$:
\begin{equation*}
U\left( 1,4\right) =0.4*u(4)>U\left( 0,4\right) =0*u(5)=0
\end{equation*}

which holds for any $u$.

\item For an equilibrium with $C=10$: 
\begin{equation*}
U\left(2,8\right)=0.8*u(3)>U\left(0,8\right)=0.4*u(5)
\end{equation*}

which holds as long as $u(5)<2*u(3)$.
\end{enumerate}

\begin{observation}\label{ob2}  
Comparing this treatment to Risk-Risk, we see that introducing ambiguity on the chances of total loss has a clear positive effect on equilibrium contributions. First, ambiguity on the chances of total loss reduces the likelihood of $C=0$ since players would be using weakly dominated strategies. Second, ambiguity on the chances of total loss makes the contribution profile with $C=5$ more likely, since it is an equilibrium for any $u$. Third,  ambiguity on the chances of total loss makes the contribution profile with $C=10$ more likely since it holds as an equilibrium under weaker conditions ($u(5)<2*u(3)$  vs  $u(5)<(9/5)*u(3)$).
\end{observation}

\subsection{Ambiguity-ambiguity}

In this treatment there is ambiguity in the abatement costs and the chances of a total loss.

Thus, under the probability distributions (a') and (b'), if we choose the lowest possible $p$ for each $c_{i}$ we get 
\begin{equation*}
p\left( C\right) =\left\{ 
\begin{array}{c}
0\text{ if }C<10 \\ 
0.8\text{ if }10\leq C<25%
\end{array}%
\right.
\end{equation*}%

\begin{enumerate}
\item For an equilibrium with $C=0$:
\begin{equation*}
U\left( 0,0\right) =0*u(5)=U\left( 5,0\right) =0*u(0)=0
\end{equation*}

so $C=0$ is an equilibrium, but it is not undominated.

\item For an equilibrium with $C=10$: 
\begin{equation*}
U\left( 2,8\right) =0.8*u(3)>U\left( 0,8\right) =0*u(5)=0
\end{equation*}

which holds for any $u$.
\end{enumerate}

\begin{observation}\label{ob3} 
{When we introduce ambiguity both on the chances of total loss and the abatement cost, the positive effect on equilibrium contributions is even more explicit. First, $C=0$ is  less likely since players would be using weakly dominated strategies. Second, $C=5$ is no longer an equilibrium. Finally, $C=10$ is an equilibrium for any $u$.}
\end{observation}

\subsection{Summary}

The following table shows the total contributions $C$ for pure strategy undominated Nash equilibria for risk neutral players under the different treatments:
\begin{center}
\begin{tabular}{lllll}
Equilibrium/Treatment & RR & RA & AR & AA \tabularnewline $C=0$ & Y & Y & 
&  \tabularnewline $C=5$ & Y &  & Y &  \tabularnewline $C=10$ & Y & Y & Y
& Y\tabularnewline%
\end{tabular}
\end{center}

If we only look at which total contributions are equilibria for all utility functions $u(x)$we are left with:

\begin{center}
\begin{tabular}{lllll}
Equilibrium/Treatment & RR & RA & AR & AA \tabularnewline $C=0$ & Y & Y & 
&  \tabularnewline $C=5$ &  &  & Y & \tabularnewline $C=10$ &  &  &  & Y%
\tabularnewline%
\end{tabular}
\end{center}

Clearly, large  total equilibrium contributions $C$ require a population with very homogeneous and narrow preferences, except for the case of ambiguity in both dimensions. Low total contributions are robust under risk, but are not equilibrium under ambiguity in the probability of the total loss. The only case for which intermediate contributions are in equilibrium robustly is with ambiguity about the probability, but not of the abatement cost. With these equilibrium characterization we can now propose our hypotheses.


\subsection{Hypotheses}\label{hypotheses}

\begin{enumerate}
\item[H1:] Contributions should be larger in AA than in the other three treatments (RR, AR and RA).

\item[H2:] Contribution should be larger in AR than in RR and in RA.

\item[H3:] It is harder to rank RR and RA but contributions could be more
polarized in RA than in RR.
\end{enumerate}

We have assumed no social preferences. If social preferences are important enough, the prediction is that ambiguity 
should not matter at all. Subjects will reach the $C=10$ equilibrium in all treatments, 
leaving no room for ambiguity to have an effect. However, with weak social preferences, the prediction that ambiguity increases contributions would still hold.

Our equilibrium analysis has assumed maxmin expected utility preferences, which is consistent with ambiguity aversion, and implies
that subjects have pessimistic beliefs under ambiguity. There are alternative assumptions that we could use, for example the maxmax
expected utility preferences, which implies optimistic beliefs and generally lower contributions under ambiguity 
(see the appendix for the predictions of the model in this case).

\section{Data and sample}

We ran a survey-experiment (conducted by ASU Research) of 1500 individuals. It was a representative sample the Spanish population at the level of regions, age, gender and education. On average subjects earned 5.16\euro. 

Table \ref{tab:sample} describes the number and percentage of subjects in the sample by region, and the corresponding values in the population (the population data is from INE, the National Statistical Institute of Spain).

 \begin{table}[htbp]
   \centering
   \caption{Representative sample.}
 \begin{adjustbox}{max width=0.9\textwidth}
     \begin{tabular}{lcccc}
     \hline
       & \multicolumn{2}{c}{Sample} & \multicolumn{2}{c}{Population (2020)*}\\
     \hline
     Andalucía & 279 & 18.6\% & 8,464,411 & 17.8\% \\
     Aragón & 45 & 3.0\% & 1,329,391 & 2.8\% \\
     Principado de Asturias & 30 & 2.0\% & 1,018,784 & 2.1\% \\
     Baleares & 25 & 1.7\% & 1,171,543 & 2.5\% \\
     Canarias & 75 & 5.0\% & 2,175,952 & 4.6\% \\
     Cantabria & 15 & 1.0\% & 582,905 & 1.2\% \\
     Castilla-La Mancha & 60 & 4.0\% & 2,045,221 & 4.3\% \\
     Castilla y León & 75 & 5.0\% & 2,394,918 & 5.0\% \\
     Cataluña & 242 & 16.1\% & 7,780,479 & 16.4\% \\
     Extremadura & 30 & 2.0\% & 1,063,987 & 2.2\% \\
     Galicia & 90 & 6.0\% & 2,701,819 & 5.7\% \\
     La Rioja & 12 & 0.8\% & 319,914 & 0.7\% \\
     Comunidad de Madrid & 217 & 14.5\% & 6,779,888 & 14.3\% \\
     Región de Murcia & 45 & 3.0\% & 1,511,251 & 3.2\% \\
     Comunidad Foral de Navarra & 20 & 1.3\% & 661,197 & 1.4\% \\
     País Vasco & 77 & 5.1\% & 2,220,504 & 4.7\% \\
     Comunidad Valenciana & 163 & 10.9\% & 5,057,353 & 10.7\% \\
     \hline
     Total & 1,500 & 100\% & 42,222,164 & 100\% \\
     \hline
     \end{tabular}%
     \end{adjustbox}
   \label{tab:sample}%
 \end{table}%

In table \ref{tab:summary} we report summary statistics of the main covariates of our study. One interesting observation is that the correlation between ambiguity and risk aversion is $-0.41$ with a p-value of 0.00.

 \begin{table}[htbp]
   \centering
   \caption{Summary statistics}
 \begin{adjustbox}{max width=0.99\textwidth}
     \begin{tabular}{lrrrrr}
     \hline
       & \multicolumn{1}{l}{Mean} & \multicolumn{1}{l}{p50} & \multicolumn{1}{l}{SD} & \multicolumn{1}{l}{Min} & \multicolumn{1}{l}{Max} \\
     \hline
     $age$ & 43.84 & 43 & 14.06 & 18 & 74 \\
     $female$ & 0.52 & 1 & 0.50 & 0 & 1 \\
     $education$ & 2.95 & 3 & 1.34 & 1 & 5 \\
     $patience$ & 3.37 & 4 & 2.17 & 0 & 6 \\
     $CRT$ & 1.59 & 2 & 0.97 & 0 & 3 \\
     $math ability$  & 2.11 & 2 & 0.87 & 0 & 3 \\
     $altruism$ & 1.64 & 2 & 0.77 & 0 & 3 \\
     $envy$  & 2.16 & 3 & 1.30 & 0 & 4 \\
     $ideology$ $(right)$ &  4.92 & 5 & 2.31 & 1 & 10 \\
     $gravity$ $(warming)$ &  7.69 & 8 & 1.78 & 1 & 10 \\
     $number actions$ &  4.59 & 5 & 2.21 & 1 & 11 \\
     $unemployed$ & 0.12 & 0 & 0.33 & 0 & 1 \\
     $social transfer$ & 0.19 & 0 & 0.39 & 0 & 1 \\
     $risk$ $aversion*$ & 0.04 & -0.1 & 0.29 & -0.1 & 1 \\
     $ambiguity$ $aversion*$ & 0.02 & 0 & 0.47 & -2 & 2 \\
     $contribution$ $PGG$ & 2.72 & 2 & 1.39 & 0 & 5 \\
          \hline
        
    Note: See the appendix for the definition of the covariates. 
     \\
     \hline
     \end{tabular}%
     \end{adjustbox}
   \label{tab:summary}%
 \end{table}%

\newpage

\subsection{Balance}

Table \ref{tab:my-table} reports the balance tests across treatments for the main covariates. Only three of the variables show 
a small significant difference, something to be expected with so many covariates.\footnote{This significant result would not survive a multiple hypothesis testing correction, like Bonferroni.} The differences are nevertheless quite small. 
So the randomization appears to be successful.

\begin{table}[htbp]
\caption{Balance across treatments}
\label{tab:my-table}
\begin{adjustbox}{max width=0.9\textwidth}
\begin{tabular}{lcccc}
\hline
            & $Mean_{RR}$ & $p(T_{AR}-T_{RR})$ & $p(T_{RA}-T_{RR})$ & $p(T_{AA}-T_{RR})$ \\
\hline
$age$             & 43.197 & 0.432   & 0.683    & 0.025** \\
$female$          & 0.526  & 0.794   & 0.683    & 0.684 \\
$education$     & 2.961  & 0.642   & 0.723    & 0.819 \\
$patience$        & 3.300  & 0.144   & 0.879    & 0.863 \\
$ambiguity$ $aversion$   & 0.019  & 0.185   & 0.706    & 0.288 \\
$risk$ $aversion$        & 0.041  & 0.513   & 0.980    & 0.434  \\
$CRT$   & 1.531  & 0.300   & 0.267    & 0.288 \\
$math$ $ability$    & 2.055  & 0.356   & 0.247    & 0.085*  \\
$altruism$    & 1.629  & 0.233   & 0.599    & 0.144  \\
$envy$    & 2.113  & 0.730   & 0.312    & 0.430  \\
$ideology$ $(right)$    & 4.939  & 0.657   & 0.994    & 0.979  \\
$gravity$ $(warming)$    & 7.721  & 0.777   & 0.668    & 0.793  \\
$number$ $actions$     & 4.464  & 0.731   & 0.166    & 0.032  \\
$unemployed$     & 0.111  & 0.524   & 0.445    & 0.736  \\
$social$ $transfer$ & 0.176  & 0.675   & 0.115    & 0.955  \\

\hline
\end{tabular}
\end{adjustbox}
\end{table}

\newpage

\section{Results}

\begin{figure}[htp]

\subfloat[Means. Grey lines correspond to the 95\% CI ]{%
  \includegraphics[clip,width=12cm]{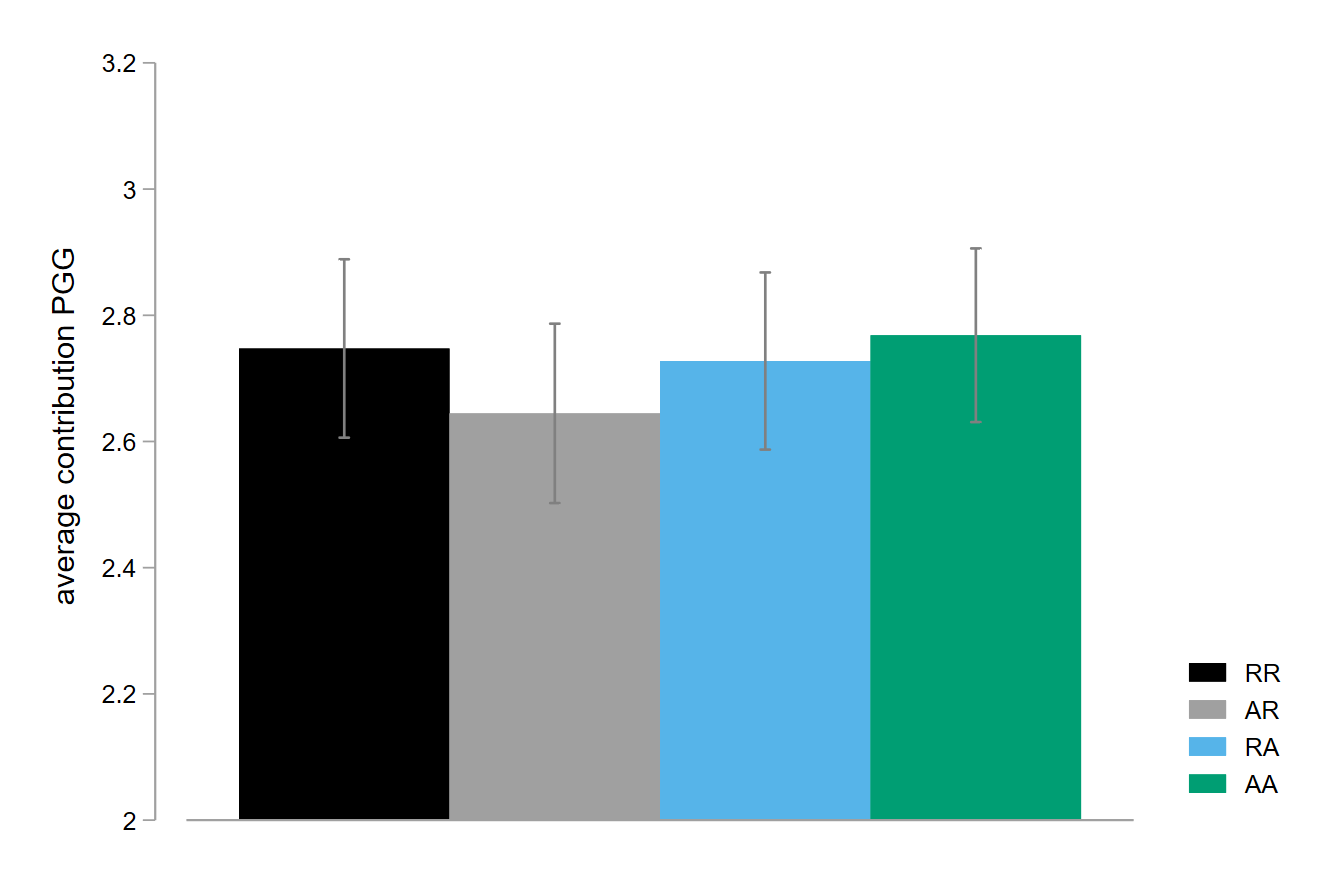}%
}

\subfloat[Histogram]{%
  \includegraphics[clip,width=12cm]{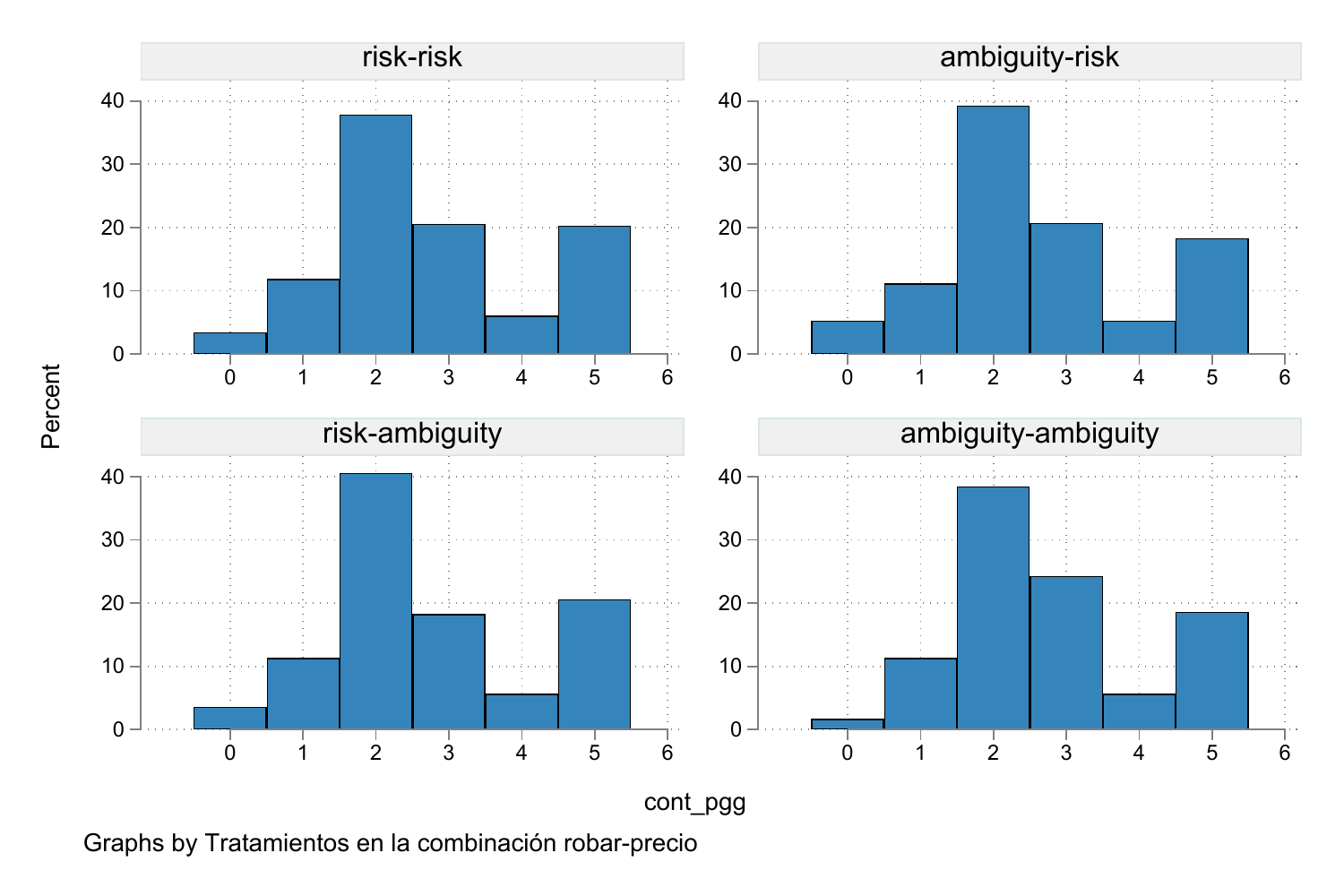}%
}

\caption{PGG contributions across treatments. }

\label{fig:ccont_pgg_bytreatra_catontributions}
\end{figure}

Our main outcome of interest is the contribution level and the comparison between treatments. An inspection of panel (a) of Figure \ref{fig:ccont_pgg_bytreatra_catontributions} 
reveals that there are no significant differences between the treatments (we also explore this via
regression analysis later).  
Ambiguity, either in the mitigation costs or the chances of total loss or both, seems to make no
difference on the amount contributed by the participants, with respect to the case of risk in both dimensions.
Beyond the average treatment effect, panel (b) of Figure \ref{fig:ccont_pgg_bytreatra_catontributions} presents the distribution of contributions by treatment, showing no differences between treatments.

Notice also that the average amount contributed by groups was above the maximum threshold to avoid total loss. 
Thus, most groups avoided the loss. 
Only 15\% of the participants contributed less than 2, and this was vastly
compensated by the 45\% of people who contributed more than 2.

\begin{figure}[htp]

\subfloat[Risk aversion]{%
  \includegraphics[width=11cm]{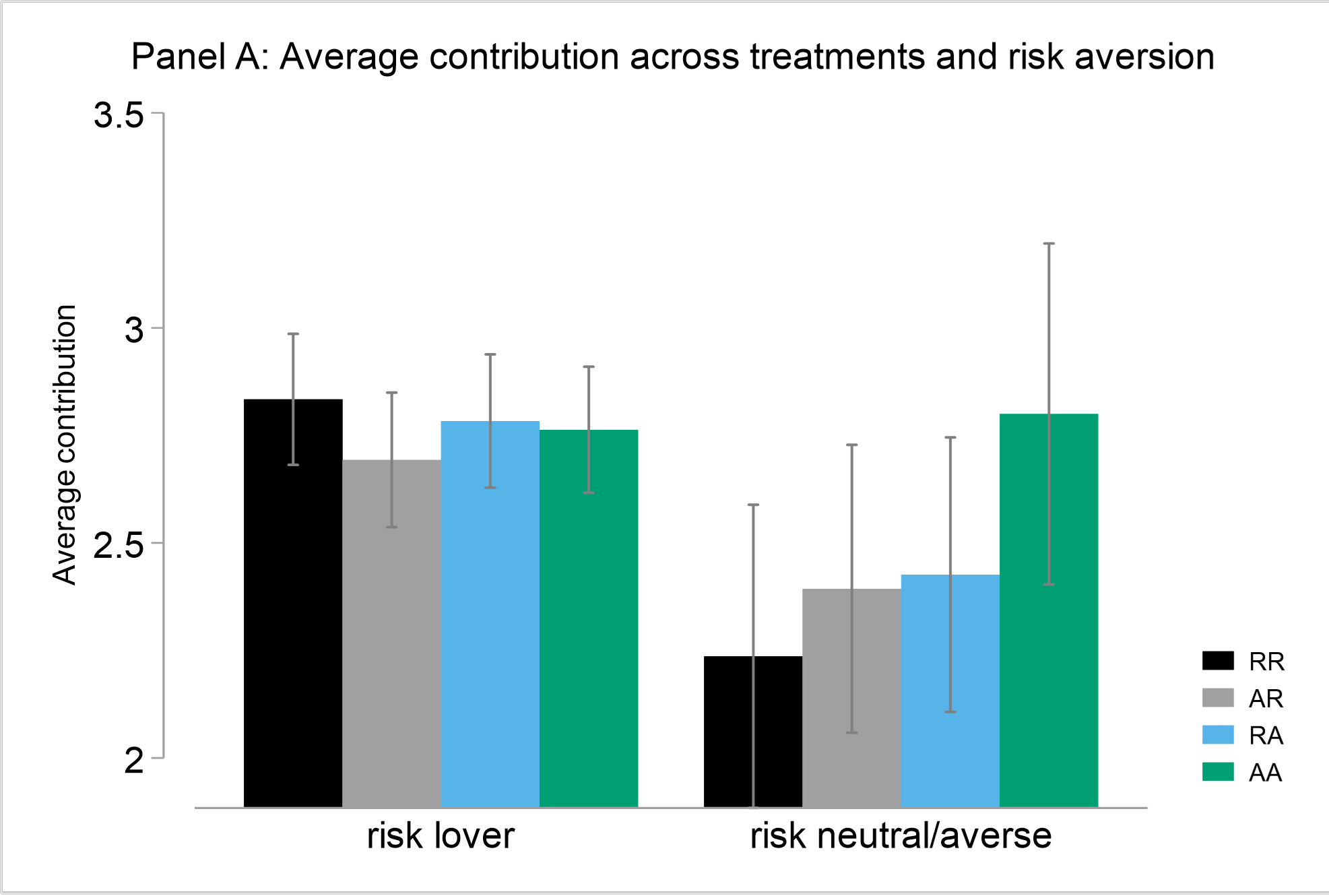}%
}

\subfloat[Ambiguity aversion]{%
  \includegraphics[width=11cm]{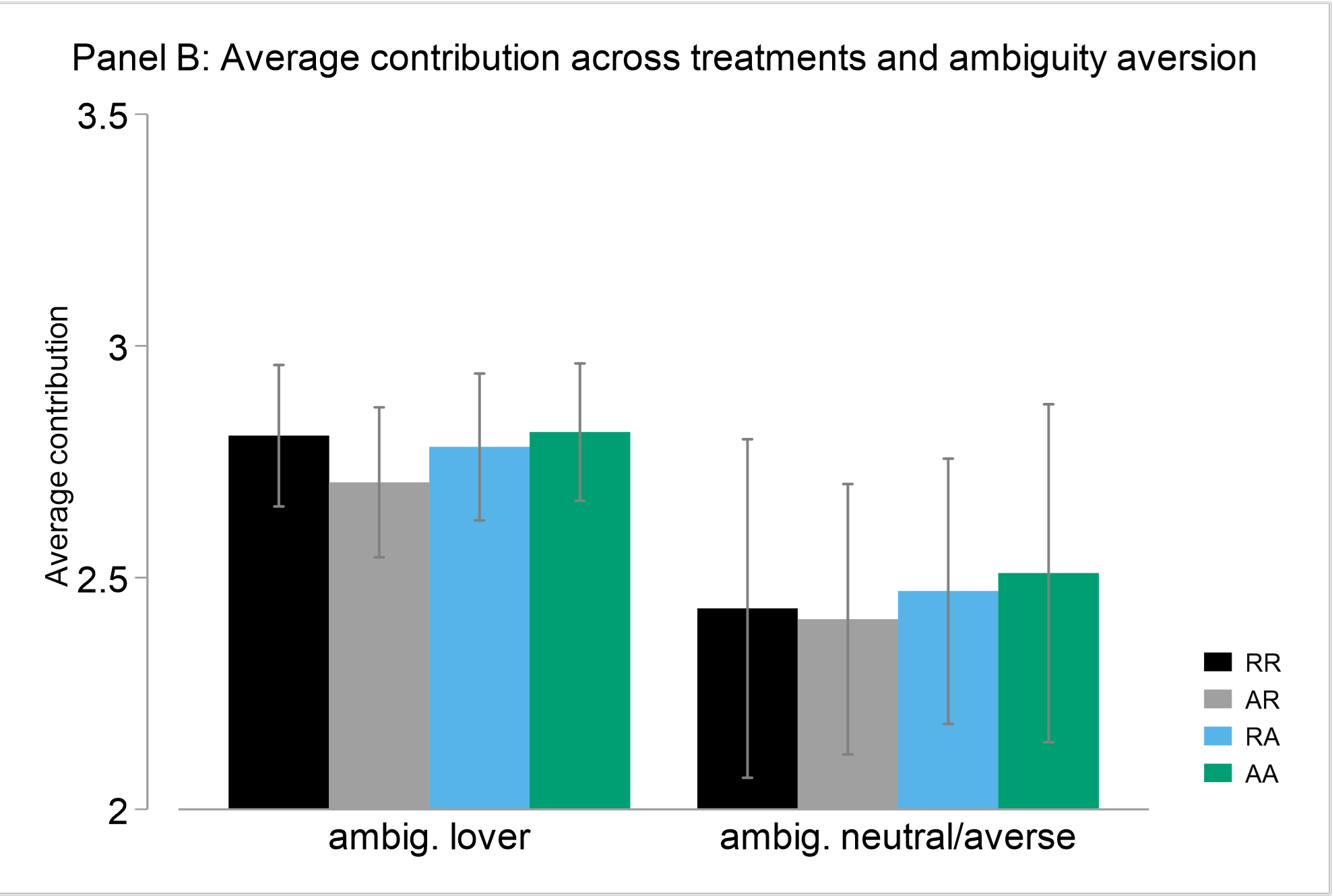}%
}

\caption{Mean contribution PGG across treatments and risk/ambiguity aversion. Grey lines are the 95\% CI. }

\label{fig:contributions-1}
\end{figure}

\newpage
    It is interesting to analyze separately the behavior of subjects with different risk or ambiguity attitudes. We split the sample between the risk loving and the risk averse participants\footnote{Risk lovers are those agents that would need to be compensated for not playing the risky lottery in the decision problem that we use to measure risk aversion.}, and between the ambiguity averse and ambiguity loving participants.\footnote{Ambiguity lovers are those agents that would need to be compensated for not playing the ambiguous lottery in the decision problem that we use to measure ambiguity aversion.} As can be seen in figure \ref{fig:contributions-1} (left panel for risk aversion, right panel for ambiguity aversion), in both cases it seems clear that risk or ambiguity averse participants contributed less on average (this will be confirmed in the regression analysis).

Table \ref{tab:Regression1-1} presents the results using OLS regressions where the dependent variable is the individual contribution in the game. Column (1) already describes the main result of the paper: none of the treatments with ambiguity yielded either higher or lower contributions than the baseline with only risk ($p>0.300$). The regression has robust standard errors and in the specification of Column (1) there are no other regressors except the constant.

The result is confirmed as we add more controls ($p>0.480$). 
{Column (2) controls for $age$, $education$, $gender$, $altruism$, $envy$, $ideology$ (higher values reflect
a right wing ideology), subjective perception about the gravity of the global warming ($gravity$), 
\emph{number of actions} that subjects did to prevent global warming, $unemployed$, whether or not they are recipients of \emph{sacial transfers}
and the number of right answers in the $CRT$ task. 
From all these variables, $age$, $education$ and the $CRT$ have a negative and significant coefficient
($p=0.043$, $p=0.035$ and $p=0.000$, respectively), while $altruism$ (marginally) increased contributions
($p<0.058$). Columns (3) and (4) show also that \emph{risk aversion}  significantly reduced the
 contribution levels ($p=0.001$ and $p=0.000$, respectively). {Additionally, Column (4) adds the ambiguity aversion variable and shows that participants who are ambiguity averse contributed less ($p=0.046$). Finally, Column (5) 
shows that the $beliefs$ about others' contributions were positively correlated with one's own contribution
(the coefficient is indeed the same in both regressions, $p=0.000$), something that makes sense 
in a game where a threshold can only be reached if others contribute enough. But the most important 
result is that \emph{risk aversion} significantly reduced contribution levels even when we
control for beliefs regardless of the treatment. Another important result, is that the magnitude of the coefficient of \emph{ambiguity aversion} is smaller when adding beliefs, but it is still negative, although now only marginally significant ($p=0.058$).}} 

{Table \ref{tab:Regression2} shows the results for the interaction of risk  and ambiguity aversion
with the treatments. As before, the treatment variables are not significant in either Column (1) or (2)
($p>0.419$). Column (1) shows that the coefficient of risk aversion is negative and 66\% larger than the one in Column (4) of Table \ref{tab:Regression1-1} ($p=0.001$). However, the interaction of this variable
with treatments RA and AA is positive and significant ($p=0.017$ and $p=0.011$). These results suggest
that risk aversion significantly reduced contribution levels for the baseline group (RR), while in
treatments RA and AA it increased contributions significantly. 
This can be confirmed by looking at Table \ref{tab:riskaverse} in the appendix, that presents the same regression with the sample restricted to risk averse individuals. Finally, Column (2) shows that \emph{ambiguity aversion} has a positive but not significant effect ($p=0.069$), while the interaction with treatment AA and  \emph{risk aversion} are negative and significant ($p=0.015$ and $p=0.001$). These results suggest, that those subjects who are ambiguity averse contributed  less under some extreme ambiguous conditions.}

 \begin{table}[htbp]
   \centering
   \caption{Regression results}
    \begin{adjustbox}{max height=0.70\textwidth}
     \begin{tabular}{lccccc}
     \hline
       & (1) & (2) & (3) & (4) & (5)\\
       & Cont. PGG & Cont. PGG & Cont. PGG & Cont. PGG & Cont. PGG  \\
     \hline
                       &           &           &           &           &           \\
$AR$                   & -0.103    & -0.066    & -0.057    & -0.054    & 0.009     \\
                       & (0.102)   & (0.100)   & (0.100)   & (0.100)   & (0.084)   \\
$RA$                   & -0.020    & 0.007     & 0.012     & 0.011     & 0.028     \\
                       & (0.102)   & (0.101)   & (0.101)   & (0.101)   & (0.082)   \\
$AA$                   & 0.021     & 0.050     & 0.047     & 0.042     & 0.060     \\
                       & (0.101)   & (0.100)   & (0.100)   & (0.100)   & (0.083)   \\
$age$                  &           & -0.005**  & -0.005**  & -0.005**  & -0.005**  \\
                       &           & (0.003)   & (0.003)   & (0.003)   & (0.002)   \\
$female$               &           & 0.012     & -0.011    & -0.026    & 0.002     \\
                       &           & (0.073)   & (0.073)   & (0.073)   & (0.059)   \\
$education$            &           & -0.060**  & -0.059**  & -0.059**  & -0.010    \\
                       &           & (0.029)   & (0.029)   & (0.029)   & (0.025)   \\
$altruism$             &           & 0.094*    & 0.105**   & 0.102**   & 0.030     \\
                       &           & (0.049)   & (0.049)   & (0.049)   & (0.040)   \\
$envious$              &           & 0.018     & 0.019     & 0.017     & 0.019     \\
                       &           & (0.027)   & (0.027)   & (0.027)   & (0.023)   \\
$idiology (right)$     &           & -0.002    & -0.000    & -0.000    & -0.001    \\
                       &           & (0.017)   & (0.017)   & (0.017)   & (0.014)   \\
$gravity (warming)$    &           & 0.039     & 0.040*    & 0.039*    & 0.005     \\
                       &           & (0.024)   & (0.024)   & (0.024)   & (0.019)   \\
$number actions$       &           & -0.015    & -0.018    & -0.017    & 0.014     \\
                       &           & (0.019)   & (0.019)   & (0.019)   & (0.015)   \\
$social transfer$      &           & 0.142     & 0.151     & 0.142     & 0.057     \\
                       &           & (0.096)   & (0.097)   & (0.097)   & (0.079)   \\
$CRT$                  &           & -0.254*** & -0.248*** & -0.250*** & -0.142*** \\
                       &           & (0.039)   & (0.039)   & (0.038)   & (0.031)   \\
$unemployed$           &           & 0.146     & 0.131     & 0.140     & 0.029     \\
                       &           & (0.120)   & (0.120)   & (0.121)   & (0.098)   \\
$risk$ $aversion$      &           &           & -0.423*** & -0.544*** & -0.337*** \\
                       &           &           & (0.130)   & (0.142)   & (0.107)   \\
$ambiguity$ $aversion$ &           &           &           & -0.179**  & -0.125*   \\
                       &           &           &           & (0.090)   & (0.066)   \\
$beliefs$              &           &           &           &           & 0.170***  \\
                       &           &           &           &           & (0.007)   \\
Constant               & 2.747***  & 3.071***  & 3.057***  & 3.081***  & 1.450***  \\
                       & (0.072)   & (0.302)   & (0.301)   & (0.301)   & (0.264)   \\
                       &           &           &           &           &           \\
Observations           & 1,500     & 1,427     & 1,421     & 1,421     & 1,421     \\
R-squared              & 0.001     & 0.060     & 0.067     & 0.070     & 0.391         \\ 
\hline
     \end{tabular}%
 \end{adjustbox}
   \label{tab:Regression1-1}%

 \begin{tablenotes}
  \scriptsize
  \item
  Note: Robust standard errors in parentheses. 
 *Beliefs refer to the total contribution expected by the others 4 members of the group. Significance levels: *** $p<0.01$, ** $p<0.05$, * $p<0.1$.
   \end{tablenotes}
 
 \end{table}

 
 \begin{table}[htbp]
  \centering
   \caption{Regression results: interaction models}
      \begin{adjustbox}{max height=0.72\textwidth}
     \begin{tabular}{lcc}
     \hline
	&	(1)	&	(2)	\\
	&	Cont. PGG	&	Cont. PGG	\\
\hline
$AR$                        & -0.003    & 0.015     \\
                            & (0.085)   & (0.084)   \\
$RA$                        & 0.001     & 0.033     \\
                            & (0.083)   & (0.082)   \\
$AA$                        & 0.039     & 0.067     \\
                            & (0.082)   & (0.083)   \\
$age$                       & -0.005**  & -0.005**  \\
                            & (0.002)   & (0.002)   \\
$CRT$                       & -0.141*** & -0.139*** \\
                            & (0.031)   & (0.031)   \\
$beliefs$                   & 0.170***  & 0.170***  \\
                            & (0.007)   & (0.007)   \\
$risk$ $aversion$           & -0.700*** & -0.339*** \\
                            & (0.208)   & (0.106)   \\
$AR$ * $risk$ $aversion$      & 0.359     &           \\
                            & (0.271)   &           \\
$RA$ * $risk$ $aversion$      & 0.679**   &           \\
                            & (0.285)   &           \\
$AA$ * $risk$ $aversion$      & 0.814**   &           \\
                            & (0.321)   &           \\
$ambiguity$ $aversion$      &           & 0.105     \\
                            &           & (0.119)   \\
$AR$ * $ambiguity$ $aversion$ &           & -0.219    \\
                            &           & (0.144)   \\
$RA$ * $ambiguity$ $aversion$ &           & -0.293    \\
                            &           & (0.182)   \\
$AA$ * $ambiguity$ $aversion$ &           & -0.434**  \\
                            &           & (0.178)   \\
Constant                    & 1.422***  & 1.434***  \\
                            & (0.263)   & (0.265)   \\
                            &           &           \\
Observations                & 1,421     & 1,421     \\
R-squared                   & 0.394     & 0.394    	\\
    \hline
     \end{tabular}%
 \end{adjustbox}
  \label{tab:Regression2}%
  
   \begin{tablenotes}
    \tiny
  \item
 Note: Robust standard errors in parentheses. Significance levels: ***$p<0.01$,**$p<0.05$, *$p<0.1$.
   \end{tablenotes}

  \end{table}

{Based on previous research we hypothesized that gender and CRT would affect contributions. In section \ref{AdAn} of the Appendix,
Table \ref{tab:othervariablesl} reports our results. Only CRT has a significant (negative) impact  on
contributions, but the interactions with the treatments are null in all cases.} We find the result on
CRT worth noting. The lack of impact of ambiguity on contributions could be due to lack of attention or comprehension of the environment/instructions by our
participants. But individuals with high CRT tend to be thoughtful and reflective (that is what CRT measures), and even for high CRT individuals, there is no significant effect of the treatment. We double checked this result repeating the analysis just for high CRT individuals (see Appendix, Table
\ref{tab:highCRT}). Furthermore, to check the possibility that the absence of effect was due to the lack of
mathematical sophistication, we also repeated the analysis in the sub-sample of participants with high scores in the
mathematical part of the questionnaire. Again, we observe no impact of the treatment (see Appendix,
table \ref{tab:mathability}).

As mentioned in the introduction, 
one interpretation of our results is that many
people, even reflective and sophisticated types, code uncertainty in rough ways. This is true even if they
are given precise estimates of the uncertainty. Thus, in the presence of either intervals or specific
values, they think of them as being similar. This is consistent with the theoretical framework (and
quantitative assessment) of \cite{khaw2021cognitive}.
    
{In order to interpret the results about risk and ambiguity aversion, we run a regression to understand
which factors are correlated with a belief that others' will contribute less (remember that a belief that
others' contribute less is associated with a lower personal contribution). In Table
\ref{tab:RegressionBeliefsAmb}, Column (1) shows that the treatments did not induce a shift in beliefs
about others' contributions ($p>0.200$). These results hold when we add controls in Column (2) ($p>190$).
Columns (3) and (4) show that \emph{risk aversion} are associated with a beliefs that
others contribute less ($p=0.018$ and $p=0.010$), while \emph{ambiguity aversion} is not significant. Finally, education, number of actions implemented to
prevent global warming and the number of reflective choices are correlated with a belief that
others contribute less in all the specifications ($p<0.003$, $p<0.007$ and $p=0.000$, respectively); while
an increase on gravity perception of the global warming and altruism increase beliefs about others'
contribution ($p<0.014$ and $p<0.013$). 

These observation can explain the somewhat puzzling fact that risk averse participants make lower
contributions. If a risk averse person is unsure about others' contributions, a best-response would be to
contribute more to avoid falling below the threshold. But we find that risk averse participants believe
that others contribute little, and they respond by lowering their own contribution. This can be seen as a
form of false consensus \citep{engelmann2000false} or projection bias
\citep{cason2014misconceptions}, which has been shown to be important in public good games like ours
\citep{smith2015modeling}. }

{Other papers have found a negative relationship between risk aversion and contributions in a public good game (\cite{teyssier2012}), although in general the relationship between risk preferences and contributions to a PGG is inconclusive (see \cite{kocher2015}).}

{Finally, we analyze the effect of strategic uncertainty. In order to do that, in Table \ref{tab:strategic} we
add three additional variables to the regression: i) subject's \textit{perception accuracy}, which measures
how reliable they think their beliefs are; ii) a dummy variable that identifies subjects who are
\textit{pivotal} (equals one if subject believes that others' contribution will be in the range [5,9) and
zero otherwise); and iii) the interaction between both variables. We find that \textit{perception accuracy} is
not significant ($p>0.100$) while the coefficients of \textit{pivotal} and the interaction of this variable with \textit{perception accuracy}
are negative and significant ($p=0.023$ and $p=0.024$, respectively)\footnote{It should be mentioned that we also run a
triple-interaction model but none of the different interactions of \textit{perception accuracy} and
\textit{pivotal} with risk aversion were significant ($p>0.300$). This is why we do not present these results.}.
These results suggest that pivotal subjects reduced their contributions and this effect is greater as they
perceive that their beliefs are more reliable.  

One could interpret this result as suggesting that strategic uncertainty is large enough so that the 
additional uncertainty about the parameters of the model does not add much ambiguity. This can complement
our alternative explanation that some people code probabilities in rough categories as in
\cite{khaw2021cognitive}. }

 \begin{table}[htbp]
   \centering
   \caption{Regression results: beliefs}
    \begin{adjustbox}{max width=0.8\textwidth}
     \begin{tabular}{lcccc}
     \hline
    	&	(1)	&	(2)	&	(3)	&	(4)	\\
	&	Beliefs 	&	Beliefs 	&	Beliefs 	&	Beliefs 	\\
	&	(cont. PGG) 	& (cont. PGG) 	& (cont. PGG) 	&	(cont. PGG) 	\\
\hline	
$AR$                   & -0.439   & -0.436    & -0.380    & -0.376    \\
                       & (0.349)  & (0.343)   & (0.343)   & (0.343)   \\
$RA$                   & -0.207   & -0.095    & -0.099    & -0.102    \\
                       & (0.357)  & (0.353)   & (0.352)   & (0.351)   \\
$AA$                   & -0.218   & -0.111    & -0.095    & -0.104    \\
                       & (0.351)  & (0.347)   & (0.349)   & (0.349)   \\
$age$                  &          & -0.003    & -0.003    & -0.003    \\
                       &          & (0.009)   & (0.009)   & (0.009)   \\
$female$               &          & -0.057    & -0.139    & -0.166    \\
                       &          & (0.253)   & (0.253)   & (0.255)   \\
$education$            &          & -0.298*** & -0.287*** & -0.286*** \\
                       &          & (0.097)   & (0.097)   & (0.097)   \\
$altruism$             &          & 0.406**   & 0.429***  & 0.423***  \\
                       &          & (0.164)   & (0.163)   & (0.164)   \\
$envious$              &          & -0.007    & -0.007    & -0.009    \\
                       &          & (0.093)   & (0.093)   & (0.093)   \\
$idiology (right)$     &          & 0.001     & 0.003     & 0.004     \\
                       &          & (0.058)   & (0.058)   & (0.058)   \\
$gravity (warming)$    &          & 0.210**   & 0.204**   & 0.203**   \\
                       &          & (0.083)   & (0.083)   & (0.083)   \\
$number actions$       &          & -0.174*** & -0.184*** & -0.184*** \\
                       &          & (0.064)   & (0.064)   & (0.064)   \\
$social transfer$      &          & 0.507     & 0.515     & 0.499     \\
                       &          & (0.326)   & (0.329)   & (0.331)   \\
$CRT$                  &          & -0.631*** & -0.633*** & -0.636*** \\
                       &          & (0.136)   & (0.137)   & (0.136)   \\
$unemployed$           &          & 0.669     & 0.641     & 0.657     \\
                       &          & (0.436)   & (0.438)   & (0.439)   \\
$risk$ $aversion$      &          &           & -1.001**  & -1.220*** \\
                       &          &           & (0.425)   & (0.472)   \\
$ambiguity$ $aversion$ &          &           &           & -0.323    \\
                       &          &           &           & (0.313)   \\
Constant               & 9.187*** & 9.480***  & 9.571***  & 9.614***  \\
                       & (0.251)  & (1.033)   & (1.029)   & (1.029)   \\
                       &          &           &           &           \\
Observations           & 1,500    & 1,427     & 1,421     & 1,421     \\
R-squared              & 0.001    & 0.058     & 0.061     & 0.062      \\
     \hline
     \end{tabular}%
      \end{adjustbox}
   \label{tab:RegressionBeliefsAmb}%
   
 \begin{tablenotes}
 \scriptsize
  \item Note: Robust standard errors in parenthesis. *Beliefs refers to the total contribution expected by the others 4 members of the group. Significance levels: ***$p<0.01$,**$p<0.05$, *$p<0.1$.
    \end{tablenotes}
    
 \end{table}

 \begin{table}[htbp]
   \centering
   \caption{Regression results: Strategic uncertainty}
    \begin{adjustbox}{max height=0.7\textwidth}
     \begin{tabular}{lc}
\hline			
	&	(1)	\\
	&	Cont. PGG	\\
\hline			
$AR$                                   & -0.036    \\
                                     & (0.099)   \\
$RA$                                   & 0.034     \\
                                     & (0.099)   \\
$AA$                                   & 0.075     \\
                                     & (0.096)   \\
$age$                                  & -0.007*** \\
                                     & (0.003)   \\
$female$                               & 0.022     \\
                                     & (0.071)   \\
$education$                            & -0.057**  \\
                                     & (0.028)   \\
$altruism$                             & 0.089*    \\
                                     & (0.049)   \\
CRT                                  & -0.204*** \\
                                     & (0.039)   \\
$pivotal$                              & -0.315**  \\
                                     & (0.139)   \\
$perception$ $accuracy$                  & 0.003     \\
                                     & (0.002)   \\
$pivotal$ $\times$ $perception$ $accuracy$ & -0.005**  \\
                                     & (0.002)   \\
Constant                             & 3.192***  \\
                                     & (0.320)     \\
                                     &           \\
Observations                         & 1,427     \\
R-squared                            & 0.108    \\
\hline			

     \hline
     \end{tabular}%

 \end{adjustbox}
   \label{tab:strategic}%

 \begin{tablenotes}
 \scriptsize
  \item Note: Robust standard errors in parenthesis. $Beliefs$ refers to the total contribution expected by the others 4 members of the group. $Pivotal$ is a dummy variable, which equals one if subject believes that others' contribution will be in the range [5,9) and zero otherwise. $Perception accuracy$ measure how reliable is his perception about others´ contributions. Significance levels: ***$p<0.01$,**$p<0.05$, *$p<0.1$.
 \end{tablenotes}
 \end{table}

\newpage
\section{Conclusions}

We have run a representative sample experiment to test the impact of ambiguity on contributions in a threshold public good game. 
Our main result is that ambiguity did not have an effect on the mean or the distribution of contributions. 
An additional result is that risk averse people contributed less. 
We believe these are important findings to the extent that the reaction to ambiguity may influence policy. 
Our data come from a representative sample of the population, and the problem we analyze mirrors the structure of substantive real-life issues like climate change.

In the previous literature on ambiguity in games, the focus was not on uncertainties concerning the environment but on the behavior of the opponents. 
Our study addresses the effect of ambiguity on the environment when there is also strategic ambiguity. 
It turns out that the additional ambiguity about the game does not have an effect when added to the strategic uncertainty already present in all treatments.
Thus, a decreasing marginal effect of additional ambiguity may be behind our results.

In our experiment, subjects contributions were large, $c_i=2.72$ on average, higher than the strategy at the symmetric high contribution equilibrium (where $C=\sum_{i=1}^n c_i=10$). 
It is possible that altruism dominated the
perception of the problem and obscured the effect of ambiguity. 
This is consistent with the possibility of a projection bias that leads participants to contribute like they believe others do. 
This behavior could also be found in real life problems with the same structure. 
Subjects may have seen the problem as a collective decision to solve the common problem of total loss and this may have prompted cooperation. 
If other-regarding preferences have increased contributions until the high-contribution equilibrium level, ambiguity cannot possibly have an additional effect. 

{In our theoretical setup ambiguity induces a pessimistic view about the chances of a total loss, which causes larger equilibrium contributions than in the risk condition. However, actual preferences may not be of the maxmin type. In fact, in the case of climate change, ambiguity about the chances of a global disaster may leave room for some people holding "too optimistic" views. The lack of effect of ambiguity in our experiment is consistent with preferences that are intermediate between the pessimist maxmin and the optimist maxmax preferences (see Appendix A.1).}

Even though in our experiment stakes are too low to adequately represent the enormous social costs of climate change, we can derive some conclusions from the experiment. 
Science will probably reduce the uncertainties surrounding the chances of a climate disaster, reducing the level of ambiguity. 
However, given the presence of strategic uncertainty and possibly social preferences, our results suggest that the reduction in ambiguity may not impact voluntary contributions. 
Thus, a tentative policy implication is that reducing uncertainty in social problems (like climate change) will not affect the level of
contributions even under ambiguity aversion.


\bibliographystyle{apalike}
\bibliography{references}

\begin{thebibliography}{}

\bibitem[Adilov et~al., 2018]{adilov2018economic}
Adilov, N., Alexander, P.~J., and Cunningham, B.~M. (2018).
\newblock An economic “kessler syndrome”: A dynamic model of earth orbit
  debris.
\newblock {\em Economics Letters}, 166:79--82.

\bibitem[Berger and Bosetti, 2020]{berger2020policymakers}
Berger, L. and Bosetti, V. (2020).
\newblock Are policymakers ambiguity averse?
\newblock {\em The Economic Journal}, 130(626):331--355.

\bibitem[Brañas-Garza et~al., 2022]{branastime}
Brañas-Garza, P., Jorrat, D., Espín, A., and Sanchez, A. (2022).
\newblock Paid and hypothetical time preferences are the same: Lab, field and
  online evidence.
\newblock {\em Experimental Economics}, in press.

\bibitem[Calford, 2021]{calford2021mixed}
Calford, E.~M. (2021).
\newblock Mixed strategies and preference for randomization in games with
  ambiguity averse agents.
\newblock {\em Journal of Economic Theory}, 197:105326.

\bibitem[Cason and Plott, 2014]{cason2014misconceptions}
Cason, T.~N. and Plott, C.~R. (2014).
\newblock Misconceptions and game form recognition: Challenges to theories of
  revealed preference and framing.
\newblock {\em Journal of Political Economy}, 122(6):1235--1270.

\bibitem[Chalioti, 2019]{chalioti2019spillover}
Chalioti, E. (2019).
\newblock Spillover feedback loops and strategic complements in r\&d.
\newblock {\em Journal of Public Economic Theory}, 21(6):1126--1142.

\bibitem[Corgnet et~al., 2015]{corgnet2015cognitive}
Corgnet, B., Esp{\'\i}n, A.~M., and Hern{\'a}n-Gonz{\'a}lez, R. (2015).
\newblock The cognitive basis of social behavior: cognitive reflection
  overrides antisocial but not always prosocial motives.
\newblock {\em Frontiers in Behavioral Neuroscience}, 9:287.

\bibitem[Dannenberg et~al., 2011]{dannenberg2011coordination}
Dannenberg, A., L{\"o}schel, A., Paolacci, G., Reif, C., and Tavoni, A. (2011).
\newblock Coordination under threshold uncertainty in a public goods game.
\newblock {\em ZEW-Centre for European Economic Research Discussion Paper},
  11-065.

\bibitem[Di~Mauro and Finocchiaro~Castro, 2011]{di2011kindness}
Di~Mauro, C. and Finocchiaro~Castro, M. (2011).
\newblock Kindness, confusion, or… ambiguity?
\newblock {\em Experimental Economics}, 14(4):611--633.

\bibitem[Eichberger and Kelsey, 2002]{eichberger2002strategic}
Eichberger, J. and Kelsey, D. (2002).
\newblock Strategic complements, substitutes, and ambiguity: the implications
  for public goods.
\newblock {\em Journal of Economic Theory}, 106(2):436--466.

\bibitem[Eichberger et~al., 2008]{eichberger2008granny}
Eichberger, J., Kelsey, D., and Schipper, B.~C. (2008).
\newblock Granny versus game theorist: Ambiguity in experimental games.
\newblock {\em Theory and Decision}, 64(2):333--362.

\bibitem[Engelmann and Strobel, 2000]{engelmann2000false}
Engelmann, D. and Strobel, M. (2000).
\newblock The false consensus effect disappears if representative information
  and monetary incentives are given.
\newblock {\em Experimental Economics}, 3(3):241--260.

\bibitem[Gilboa et~al., 1989]{gilboa1989maxmin}
Gilboa, I., Schmeidler, D., et~al. (1989).
\newblock Maxmin expected utility with non-unique prior.
\newblock {\em Journal of Mathematical Economics}, 18(2):141--153.

\bibitem[Isaac et~al., 1989]{isaac1989assurance}
Isaac, R.~M., Schmidtz, D., and Walker, J.~M. (1989).
\newblock The assurance problem in a laboratory market.
\newblock {\em Public Choice}, 62(3):217--236.

\bibitem[Ivanov, 2011]{ivanov2011attitudes}
Ivanov, A. (2011).
\newblock Attitudes to ambiguity in one-shot normal-form games: An experimental
  study.
\newblock {\em Games and Economic Behavior}, 71(2):366--394.

\bibitem[Kelsey and Le~Roux, 2015]{kelsey2015experimental}
Kelsey, D. and Le~Roux, S. (2015).
\newblock An experimental study on the effect of ambiguity in a coordination
  game.
\newblock {\em Theory and Decision}, 79(4):667--688.

\bibitem[Kelsey and Le~Roux, 2017]{kelsey2017dragon}
Kelsey, D. and Le~Roux, S. (2017).
\newblock Dragon slaying with ambiguity: theory and experiments.
\newblock {\em Journal of Public Economic Theory}, 19(1):178--197.

\bibitem[Kelsey and le~Roux, 2018]{kelsey2018strategic}
Kelsey, D. and le~Roux, S. (2018).
\newblock Strategic ambiguity and decision-making: an experimental study.
\newblock {\em Theory and Decision}, 84(3):387--404.

\bibitem[Khaw et~al., 2021]{khaw2021cognitive}
Khaw, M.~W., Li, Z., and Woodford, M. (2021).
\newblock Cognitive imprecision and small-stakes risk aversion.
\newblock {\em The Review of Economic Studies}, 88(4):1979--2013.

\bibitem[Kocher et~al., 2015]{kocher2015}
Kocher, M.~G., Martinsson, P., Matzat, D., and Wollbrant, C. (2015).
\newblock The role of beliefs, trust, and risk in contributions to a public
  good.
\newblock {\em Journal of Economic Psychology}, 51:236--244.

\bibitem[Lenton, 2011]{lenton2011early}
Lenton, T.~M. (2011).
\newblock Early warning of climate tipping points.
\newblock {\em Nature Climate Change}, 1(4):201--209.

\bibitem[Lontzek et~al., 2015]{lontzek2015stochastic}
Lontzek, T.~S., Cai, Y., Judd, K.~L., and Lenton, T.~M. (2015).
\newblock Stochastic integrated assessment of climate tipping points indicates
  the need for strict climate policy.
\newblock {\em Nature Climate Change}, 5(5):441--444.

\bibitem[Maccheroni et~al., 2006]{maccheroni2006ambiguity}
Maccheroni, F., Marinacci, M., and Rustichini, A. (2006).
\newblock Ambiguity aversion, robustness, and the variational representation of
  preferences.
\newblock {\em Econometrica}, 74(6):1447--1498.

\bibitem[Machina and Siniscalchi, 2014]{machina2014ambiguity}
Machina, M.~J. and Siniscalchi, M. (2014).
\newblock Ambiguity and ambiguity aversion.
\newblock In {\em Handbook of the Economics of Risk and Uncertainty}, volume~1,
  pages 729--807. Elsevier.

\bibitem[Quiggin, 2005]{quiggin2005precautionary}
Quiggin, J. (2005).
\newblock The precautionary principle in environmental policy and the theory of
  choice under uncertainty.
\newblock Technical report, University of Queensland.

\bibitem[Rondeau et~al., 1999]{rondeau1999voluntary}
Rondeau, D., Schulze, W.~D., and Poe, G.~L. (1999).
\newblock Voluntary revelation of the demand for public goods using a provision
  point mechanism.
\newblock {\em Journal of Public Economics}, 72(3):455--470.

\bibitem[Smith, 2015]{smith2015modeling}
Smith, A. (2015).
\newblock Modeling the dynamics of contributions and beliefs in repeated public
  good games.
\newblock {\em Economics Bulletin}, 35(3):1501--1509.

\bibitem[Teyssier, 2012]{teyssier2012}
Teyssier, S. (2012).
\newblock Inequity and risk aversion in sequential public good games.
\newblock {\em Public Choice}, 151(1/2):91--119.

\end{thebibliography}


\newpage

\appendix
\renewcommand\thefigure{S\arabic{figure}}
\setcounter{figure}{0}
\renewcommand\thetable{S\arabic{table}}
\setcounter{table}{0}

\section{Appendix}

\subsection{Maxmax preferences}
Under the maxmax expected utility assumption, p(C) in the risk-ambiguity treatment is:
\begin{equation*}
p\left( C\right) =\left\{ 
\begin{array}{c}
0.1\text{ if }C<5 \\ 
0.9\text{ if }5\leq C<25%
\end{array}%
\right.
\end{equation*}

In the ambiguity-risk treatment:

\begin{equation*}
p\left( C\right) =\left\{ 
\begin{array}{c}
0.2\text{ if }C<5 \\ 
0.2\ast \frac{1}{2}+1\ast \frac{1}{2}=0.6\text{ if }5\leq C<10 \\ 
1\text{ if }10\leq C<25%
\end{array}%
\right.
\end{equation*}%

and in the ambiguity-ambiguity treatment:

\begin{equation*}
p\left( C\right) =\left\{ 
\begin{array}{c}
0.2\text{ if }C<5 \\ 
1\text{ if }5\leq C<25%
\end{array}%
\right.
\end{equation*}

The following table shows the pure strategy undominated Nash equilibria for risk neutral players:
\begin{center}
\begin{tabular}{lllll}
Equilibrium/Treatment & RR & RA & AR & AA \tabularnewline $C=0$ & Y & Y & Y
& Y \tabularnewline $C=5$ & Y & Y & Y & Y \tabularnewline $C=10$ & Y &  & 
& \tabularnewline%
\end{tabular}
\end{center}

If we restrict equilibria to those robust to any utility function $u(x)$:

\begin{center}
\begin{tabular}{lllll}
Equilibrium/Treatment & RR & RA & AR & AA \tabularnewline $C=0$ & Y & Y & Y
& Y \tabularnewline $C=5$ &  &  &  & \tabularnewline $C=10$ &  &  &  & %
\tabularnewline%
\end{tabular}
\end{center}

Note that the optimism inherent to the maxmax preferences translates into lower equilibrium contributions.

\newpage

\subsection{Covariates}\label{Cov}

\begin{itemize}
\item \textit{age}: age of the participant.

\item \textit{female}: 1 if female, 0 otherwise.

\item \textit{education}: highest level of education, 1 "Secondary school or lower", 2 "Intermediate vocational training or Baccalaureate", 3 "Higher level vocational training", 4 "Undergraduate or graduate" and 5 "Master or PhD"

\item \textit{patience}: number of future allocations in the Multi Price List tasks (6 decisions).

\item \textit{CRT}: number of correct answers in the Cognitive Reflection Test (3 questions).

\item \textit{math ability}: number of correct answers in 3 questions regarding percentages and simple interest rate. 

\item \textit{altruism}: number of altruistic options chosen in 3 of the mini Dictator Games (decisions 1, 2 and 4). In these decisions the subject is willing to give up something in exchange for making others better off.

\item \textit{envy}: number of altruistic options chosen in 3 of the mini Dictator Games (decisions 3, 5 and 6). In these decisions the subject is willing to give up something in exchange for others to be worse off.

\item \textit{ideology (right)}: self-reported by the subject on a scale from 1 (extreme left) to 10 (extreme right).

\item \textit{gravity (warming)}: self-reported subjective perception about the gravity of global warming on a scale from 1 (very low severity) to 10 (extremely severe).

\item \textit{number actions}: the number of actions aimed at combating climate change that the subject has personally taken; from a list of eleven actions, the number of actions that the subject reports to have undertaken.

\item \textit{unemployed}: self-reported by the subject; 1 if unemployed, 0 otherwise.

\item \textit{social transfer}: 1 if the subject is a recipient of social help given by central, state or local institutional units; 0 otherwise.

\item \textit{risk aversion}: risk aversion measure based on whether or not subjects decide to participate in the lottery or sell it for the price that they select. The measure ranges from -1 to 1, where high values represents a higher risk aversion. 

\item \textit{ambiguity aversion}: ambiguity aversion measure based on whether or not subjects decide to participate in the ambiguous lottery or sell it for the price that they select. The measure ranges from -1 to 1, where high values represents a higher ambiguity aversion. 

\end{itemize}

\subsection{Additional analysis}\label{AdAn}

 \begin{table}[h]
\centering
   \caption{Other interactions models}
    \begin{adjustbox}{max width=0.62\smallestside}
     \begin{tabular}{lcc}
\hline					

	& (1)	&	(2) 	\\	
	& 	Contributions PGG	&	 Contributions PGG	\\
\hline					
	&		&		\\
AR            & 0.054     & 0.048    \\
              & (0.118)   & (0.179)  \\
RA            & 0.118     & -0.027   \\
              & (0.114)   & (0.177)  \\
AA            & 0.019     & 0.163    \\
              & (0.126)   & (0.176)  \\
age           & -0.005**  & -0.005** \\
              & (0.002)   & (0.002)  \\
female        & 0.059     & 0.008    \\
              & (0.125)   & (0.060)  \\
CRT           & -0.137*** & -0.130** \\
              & (0.031)   & (0.060)  \\
beliefs       & 0.170***  & 0.170*** \\
              & (0.007)   & (0.007)  \\
risk aversion & -0.255**  & -0.254** \\
              & (0.102)   & (0.102)  \\
AR*female     & -0.085    &          \\
              & (0.168)   &          \\
RA*female     & -0.173    &          \\
              & (0.164)   &          \\
AA*female     & 0.085     &          \\
              & (0.168)   &          \\
AR*CRT        &           & -0.025   \\
              &           & (0.085)  \\
RA*CRT        &           & 0.035    \\
              &           & (0.084)  \\
AA*CRT        &           & -0.061   \\
              &           & (0.084)  \\
Constant      & 1.381***  & 1.416*** \\
              & (0.273)   & (0.270)  \\
              &           &          \\
Observations  & 1,421     & 1,421    \\
R-squared     & 0.391     & 0.390 	\\

    \hline
     \end{tabular}%
 \end{adjustbox}
    \label{tab:othervariablesl}%

Note: Robust standard errors in parenthesis. Significance levels: ***$p<0.01$,**$p<0.05$, *$p<0.1$.
  \end{table}

 \begin{table}[htbp]
   \centering
   \caption{High CRT individuals}
     \begin{tabular}{lcc}
\hline
 & (1) & (2) \\
VARIABLES & Cont. PGG & Cont. PGG\\ \hline
 &  &  \\
AR            & 0.012    & -0.048   \\
              & (0.155)  & (0.153)  \\
RA            & 0.099    & 0.028    \\
              & (0.159)  & (0.159)  \\
AA            & 0.337**  & 0.221    \\
              & (0.165)  & (0.168)  \\
risk aversion & -0.103   & -0.160   \\
              & (0.179)  & (0.186)  \\
Constant      & 2.021*** & 2.301*** \\
              & (0.118)  & (0.570)  \\
              &          &          \\
Observations  & 296      & 285      \\
R-squared     & 0.020    & 0.089    \\
Controls      & No       & Yes      \\
\hline
\end{tabular} 
   \label{tab:highCRT}%

 Note: Robust standard errors in parenthesis. Significance levels: ***$p<0.01$,**$p<0.05$, *$p<0.1$.
\end{table}    

\begin{table}[htbp]
   \centering
   \caption{Regression results for risk averse}
     \begin{adjustbox}{max height=0.53\textwidth}
\begin{tabular}{lccc} 
\hline
              & (1)       & (2)       & (3)       \\
              & Cont. PGG & Cont. PGG & Cont. PGG \\
\hline
              &           &           &           \\
AR            & -0.057    & -0.018    & 0.206     \\
              & (0.203)   & (0.207)   & (0.171)   \\
RA            & 0.195     & 0.322     & 0.456***  \\
              & (0.213)   & (0.210)   & (0.171)   \\
AA            & 0.425*    & 0.569**   & 0.581***  \\
              & (0.226)   & (0.229)   & (0.207)   \\
risk aversion & 0.276     & 0.239     & 0.421     \\
              & (0.374)   & (0.372)   & (0.297)   \\
age           &           & -0.017*** & -0.015*** \\
              &           & (0.005)   & (0.005)   \\
CRT           &           & -0.256*** & -0.096    \\
              &           & (0.086)   & (0.066)   \\
beliefs       &           &           & 0.160***  \\
              &           &           & (0.021)   \\
Constant      & 2.140***  & 3.444***  & 1.237*    \\
              & (0.244)   & (0.743)   & (0.722)   \\
              &           &           &           \\
Observations  & 301       & 290       & 290       \\
R-squared     & 0.021     & 0.107     & 0.349     \\
Controls      & No        & Yes       & Yes         \\
\hline
\end{tabular} 
 \end{adjustbox}

 Note: Robust standard errors in parenthesis. Significance levels: ***$p<0.01$,**$p<0.05$, *$p<0.1$.
 \label{tab:riskaverse}%
\end{table}    

\begin{table}[htbp]
   \centering
   \caption{Regression results for different mathematical abilities}
    \begin{adjustbox}{max height=0.53\textwidth}
\begin{tabular}{lcc} 
\hline
              & (1)       & (2)       \\
              & Low       & High      \\
VARIABLES     & Cont. PGG & Cont. PGG \\
\hline
              &           &           \\
$AR$            & 0.014     & -0.168    \\
              & (0.146)   & (0.124)   \\
$RA$            & 0.011     & 0.033     \\
              & (0.146)   & (0.132)   \\
$AA$            & 0.070     & -0.001     \\
              & (0.146)   & (0.131)   \\
\emph{risk aversion} & -0.629**  & -0.466**  \\
              & (0.211)   & (0.184)   \\
Constant      & 2.516***  & 2.310***  \\
              & (0.393)   & (0.381)   \\
              &           &           \\
Observations  & 808       & 619       \\
R-squared     & 0.025     & 0.058     \\
Controls      & Yes       & Yes      \\      
\hline
\end{tabular} 
 \end{adjustbox}
 \label{tab:mathability}%

 Note: Robust standard errors in parenthesis. Significance levels: ***$p<0.01$,**$p<0.05$, *$p<0.1$.
\end{table}

\end{document}